\documentclass[aps,prx,twocolumn,showpacs,superscriptaddress,10pt]{revtex4-2}  
\usepackage{bbm}
\usepackage{epsfig}
\usepackage{epstopdf}
\usepackage{graphicx}
\usepackage{amsmath,amssymb}
\usepackage{amsmath,bm}

\usepackage{txfonts}
\usepackage{physics}
\usepackage{comment}
\usepackage{xcolor}
\usepackage{lipsum}
\usepackage[caption=false]{subfig}
\usepackage{lineno,blindtext}
\usepackage{diagbox, eqparbox, hhline}
\usepackage{soul}
\usepackage{xcolor}

\usepackage[normalem]{ulem}
\newcolumntype{P}[1]{>{\centering\arraybackslash}p{#1}}

\usepackage[colorlinks,citecolor=blue,linkcolor=blue,urlcolor=blue,]{hyperref}

\begin{document}
	
	\title{Magnon Nesting in Driven Two-Dimensional Quantum Magnets}

	\author{Hossein Hosseinabadi}
	\email{hhossein@uni-mainz.de}
	\affiliation{Institut f\"{u}r Physik, Johannes Gutenberg Universit\"{a}t Mainz, D-55099 Mainz, Germany}
	
	\author{Yaroslav Tserkovnyak}
	\affiliation{Department of Physics and Astronomy and Mani L. Bhaumik Institute for Theoretical Physics, University of California, Los Angeles, California 90095, USA}
	
	\author{Eugene Demler}
	\affiliation{Institute for Theoretical Physics, ETH Zurich, 8093 Zurich, Switzerland.}
	
	\author{Jamir Marino}
	\affiliation{Institut f\"{u}r Physik, Johannes Gutenberg Universit\"{a}t Mainz, D-55099 Mainz, Germany}
        \affiliation{Department of Physics, The State University of New York at Buffalo, NY 14260, USA}
	
	\begin{abstract}
		
		We uncover a new class of dynamical quantum instability in driven magnets  leading to emergent enhancement of antiferromagnetic correlations  even for purely ferromagnetic microscopic couplings. A primary parametric amplification creates a frequency-tuned nested magnon distribution in momentum space, which seeds a secondary instability marked by the emergence of enhanced antiferromagnetic correlations, mirroring the instability of nested Fermi surfaces in electronic systems. In sharp contrast to the fermionic case, however, the magnon-driven instability is intrinsically non-equilibrium and fundamentally inaccessible in thermal physics. Its quantum mechanical origin sets it apart from classical instabilities such as Faraday and modulation instabilities, which underlie several instances of   dynamical behavior observed in magnetic and cold-atom systems. 
	\end{abstract}
	
	\maketitle

	\section{Introduction}
	\label{sec:intro}

    Experimental progress in recent decades has sparked great interest in the non-equilibrium behavior of quantum many-body systems across a variety of platforms, ranging from driven quantum materials to quantum optical systems such as cold atoms and superconducting circuits~\cite{bukov2015universal,Bao_LightInduced2022,bloch2022strongly,Bloch_coldatoms2008,Blatt_trapped2012,Browaeys_Rydbergs2020,Kaufman_tweezer2021}.
    In solid state physics, pump-probe experiments and advances in time-resolved spectroscopy have enabled the study of ultrafast dynamics, shedding new light on the behavior of collective modes~\cite{Demidov_magnonBEC2007,Serga_MagnonBEC2014,Lu_2DSpec2017,Zhou_magnon_scattering2021,Shan_parametric2024,Zhang_magnon2Dspec2024} and the manipulation of order parameters, including the suppression or enhancement of charge-density-wave~\cite{tomeljak2009dynamics,Huber_CDW2014,Teitelbaum_CDW2018}, superconducting~\cite{Demsar_SC2003,Kusar_SC2008,fausti2011light,Hu_photoinduced2014,Mankowsky_lightinduced2014,Mitrano_lightinduced2016,Rowe_enhancement2023}, and magnetic orders~\cite{Disa_photoFM2023,Zheng_chiral2025} under non-equilibrium conditions.
   Cold atom experiments, while operating in a very different parameter regime, provide a complementary and highly controllable setting to explore similar fundamental questions about non-equilibrium dynamics. In particular, the response of atomic condensates to resonant driving has revealed a range of phenomena, including the formation of quantum-correlated matter-wave jets~\cite{clark_jets2017}, parametric heating of superfluids~\cite{Boulier_parametricHeating2019}, staggered superfluid states~\cite{Song_staggeredSF2022}, and the emergence of spatial patterns in driven condensates~\cite{Fujii_paternformation2024,Liebster_pattern2025}.
    Together, these advances demonstrate the broad relevance of non-equilibrium physics and provide complementary insights into the emergence of collective order far from equilibrium.

	\begin{figure}[!b]
		\centering
		\includegraphics[width=0.99\linewidth]{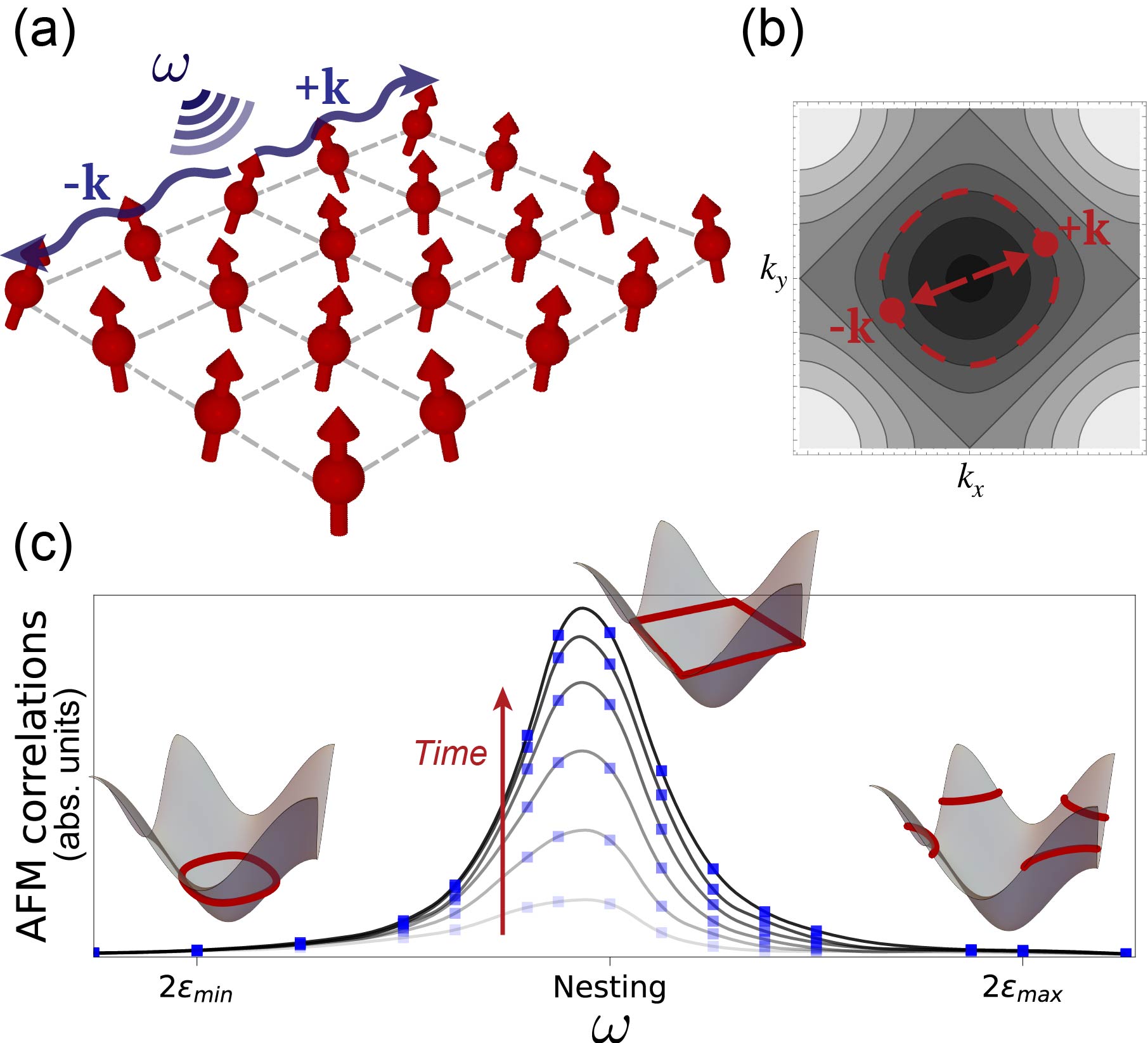}
		\caption{\textbf{Secondary instability of nested magnons by resonant parametric driving.} (a) Schematics of the setup considered in this work. An external parametric drive with frequency $\omega$ creates pair of magnons with opposite momenta. (b) Resonant drive triggers a primary instability manifested by the coherent generation of magnon pairs at resonant energies, corresponding to closed contours   in   momentum space (shown by dashes). (c) The primary instability can trigger a secondary instability, characterized by strong AFM correlations, if the driving frequency is adjusted such that the magnon distribution in the momentum space forms a nested shape, as depicted by the excitation profile in the middle.}
		\label{fig:cartoon}
	\end{figure}

    Instead, they can arise through a variety of mechanisms~\cite{delatorre_2021colloquium}, including Floquet dressing, where high-frequency driving modifies system dynamics via an effective static Hamiltonian~\cite{abanin2017rigorous,Ho_floquet2023,Hou_Goldstone2025}, and non-equilibrium manipulation of competing orders, where external driving suppresses one order to allow another to emerge~\cite{Hu_photoinduced2014,Nicoletti_photoinduced2014,Raines_enhancement2015,Sentef_competing2017,Sun_transientOrder2020}. Another prominent mechanism is resonantly driven order, in which an external drive amplifies collective modes by matching their frequencies, leading to exponential growth through parametric instabilities~\cite{cartella,PhysRevLett.84.1547,PhysRevLett.85.3680,sugiura2022resonantly,kaplan2025optically,kaplan2025spatiotemporal}. These instabilities are typically bounded by nonlinearities, but under certain conditions, they can seed secondary instabilities that generate new emergent phenomena.
    
     In classical systems, secondary instabilities are ubiquitous in fluid dynamics, such as the transition from convection to turbulence (Rayleigh-B\'{e}nard instability)~\cite{Cross_patternFormation1993} as well as the behavior of surface gravity waves (modulation instability)~\cite{Benjamin_Feir_1967,ZAKHAROV2009,zakharov2012kolmogorov}.
Secondary instabilities in quantum systems have also been theoretically studied, for instance in driven systems of bosons~\cite{Fujii_paternformation2024,zelle2025nonequilibrium} and phonons~\cite{Kaplan_PhononPattern2025}, where the formation of stable spatial patterns has been predicted. Although both examples involve quantum mechanical systems, their dynamics can largely be captured using the language of parametrically driven classical systems. An intriguing candidate for   secondary instability with genuine quantum mechanical character is photo-induced superconductivity~\cite{Knap_dynamicaCooper2016,Babadi_dynamicPairing2017,Kennes_2017,Eckhardt_resonant2024}, leading to superconductivity at temperatures well above the equilibrium critical temperature~\cite{Mitrano_lightinduced2016,Rowe_enhancement2023}. In this case, phonon-mediated interactions among electrons are strengthened by resonant driving of phonon modes, leading to stronger superconducting fluctuations. Another instance of a nonlinearity-driven quantum mechanical instability was proposed in Ref.~\cite{Neuenhahn_SineGordon2012} in the context of quantum sine-Gordon models relevant to one-dimensional Bose condensates. In this case, the primary oscillations are not triggered by external driving but arise from initial seeds provided by quantum fluctuations, which grow through nonlinear interactions and transiently form spatial patterns. Identifying further instances of genuinely non-equilibrium quantum mechanical instabilities would provide important opportunities to deepen our understanding of non-equilibrium many-body physics beyond classical analogs. 

    In this work, we propose a new mechanism of  non-equilibrium instability in two-dimensional spin systems. Inspired by the magnon-pumping experiments of Refs.~\cite{Demidov_magnonBEC2007,Serga_MagnonBEC2014,Zhou_magnon_scattering2021,Shan_parametric2024}, our approach leverages resonant driving of spins at frequency $\omega$, which induces a primary instability in the system corresponding to the rapid proliferation of magnon pairs that satisfy the energy conservation condition $\omega = \epsilon({\boldsymbol{k}}) + \epsilon(\boldsymbol{-k})$, as shown in Fig.~\ref{fig:cartoon}a. Subsequently, the growing magnon population forms an isoenergetic contour in the momentum space whose shape is determined by the driving frequency (Fig.~\ref{fig:cartoon}b). While the primary instability has a straightforward description in terms free magnons, we show that it can trigger a quantum mechanical secondary instability which crucially depends on magnon interactions. The secondary instability, characterized by a pronounced enhancement of longitudinal spin correlations at ${\boldsymbol{q}}=(\pi,\pi)$, emerges when the driving frequency $\omega$ is tuned such that the resonantly created magnons form a nested contour in momentum space, meaning a macroscopic number of points on the contour are connected to each other by the same momentum. 
    As shown in Fig.~\ref{fig:cartoon}c, we observe a peak in longitudinal spin correlations at the expected wave-vector, whose height dramatically increases as $\omega$ approaches the nesting point. Notably, \emph{the antiferromagnetic enhancement emerges even for purely ferromagnetic couplings in the system}. As stated before, the secondary instability builds upon the interactions of magnons and cannot be described within a free-particle picture. Instead, it bears a strong resemblance to the enhanced susceptibilities fermionic systems towards spin-density-waves or charge-density-waves on square lattices near half-filling~\cite{fradkin2013field,Sachdev_2023}, where the Fermi surface exhibits a similarly nested structure.  
    We emphasize that, unlike fermions, magnons obey bosonic statistics and cannot form a nested ``magnon surface" at thermal equilibrium~\cite{Girvin_Yang_2019}. Thus, the nesting-induced instability we describe is inherently a non-equilibrium phenomenon, lying beyond the conventional framework of equilibrium many-body physics. 

   This  instability differs from secondary classical instabilities such as the modulation instabilities~\cite{Benjamin_Feir_1967,agrawal2000nonlinear,ZAKHAROV2009,zakharov2012kolmogorov}, which are driven by the resonant scattering of primary modes into a set of secondary modes with both momentum and energy conserved (see~\cite{Fujii_paternformation2024} for a connection between recent experiments on driven superfluids and modulation instabilities). The secondary instability in this work originates from the inclusion of higher-order magnon processes, similar to generalized Stoner instabilities~\cite{kubler2021theory}, marking a difference with more conventional mechanisms for dynamical instabilities in many-body systems. We should mention the work in Ref.~\cite{Wang_exoticorder2020}, which considered a parametrically driven system of dissipative bosons, including the regime of driving bosons towards nesting. However, they did not observe the nesting-induced instability of the type reported in this work.

	\section{Model}
	 We consider the quantum XXZ model on a two-dimensional square lattice:
	\begin{equation}\label{H_XXZ}
		H = h \sum_i S^z_i - \frac{J}{4S}\sum_{\expval{ij}}\Big( S^x_i S^x_j + S^y_i S^y_j + \Delta \,S^z_i S^z_j \Big),
	\end{equation}
	where $S^\alpha_i$ are spin operators satisfying $su(2)$ algebra $[S^\alpha_i,S^\beta_j]=i \delta_{ij}\sum_\gamma\epsilon_{\alpha\beta\gamma}S^\gamma_i$ and $ \sum_\alpha (S^\alpha_i )^2= S(S+1)$. The couplings $h$ and $J$ are respectively, the magnetic field and the nearest-neighbor couplings. $\Delta$ is an anisotropy which can be positive or negative. For $\Delta>0$ and $\Delta<0$, the coupling between the $z$-components of spins is ferromagnetic (FM) and antiferromagnetic (AFM), respectively.
    The scaling of the couplings with $S$ is introduced to make the spin size a control parameter for the approximations that we will use later to address non-equilibrium dynamics of the system. We assume that $h$ is sufficiently large, such that the groundstate is given by the fully polarized state
	\begin{equation}\label{psi0}
		\ket{\psi_0} = \ket{\downarrow}\ket{\downarrow}\dots \ket{\downarrow}.
	\end{equation}
	
	The energy spectrum of low-lying excitations can be obtained by mapping spins to bosons using the Holstein-Primakoff transformation~\cite{Stancil_2009spinwaves}
	\begin{equation}\label{HP}
		S^z_i = a^\dagger_i a_i - S, \quad S^+_i = a^\dagger_i \sqrt{2S-a^\dagger_i a_i},
	\end{equation}
	where $S^{\pm}_i=S^x_i\pm i S^y_i$. {Expanding $H$ to the quartic order in bosonic operators yields
    \begin{equation}\label{eq:H_hp}
    H \approx H_2 + H_4,
    \end{equation}
    where the quadratic part is given by
    \begin{equation}\label{eq:H2}
        H_2 = \sum_{{\boldsymbol{k}}} \epsilon({\boldsymbol{k}})a^\dagger_{\boldsymbol{k}}a_{\boldsymbol{k}},
    \end{equation}
    with $a_{\bm{k}}=\sum_{j}e^{i\bm{k}\cdot \bm{j}}a_i/\sqrt{N}$ and $\epsilon({\boldsymbol{k}})$ the dispersion of paramagnons
	\begin{equation}\label{dispersion}
		\epsilon({\boldsymbol{k}})=h + \frac{J}{2} \,\big(2\Delta - \cos{k_x} -  \cos{k_y}\big),
	\end{equation}
    and the interaction part consisting of two contributions
    \begin{equation}\label{eq:H4}
        H_4 = H_{4}^{xy} + H_{4}^z ,
    \end{equation}
    \begin{equation}\label{eq:H4_xy}
        H_4^{xy} = \frac{1}{2N}\sum_{\bm{k,p,q}} \Big(J(\bm{k})+J(\bm{p})\Big) \,a^\dagger_{\bm{k}}a_{\bm{k-q}}a^\dagger_{\bm{p-q}}a_{\bm{p}},
    \end{equation}
    \begin{equation}\label{eq:H4_z}
        H_4^z = -\frac{\Delta}{2N}\sum_{\bm{k,p,q}} J(\bm{q}) \,a^\dagger_{\bm{k}}a_{\bm{k-q}}a^\dagger_{\bm{p-q}}a_{\bm{p}},
    \end{equation}
    where
    \begin{equation}\label{Jz_q}
		J({\boldsymbol{q}})=\frac{J}{2S} \big( \cos{q_x} + \cos{q_y} \big).
	\end{equation}
	Eq.~\eqref{eq:H_hp} describes a system of bosons with nearest-neighbor hopping on a square lattice, subject to the interactions in $H_4$. As we will discuss later in Section~\ref{sec:Bose_stoner}, $H^z_4$ plays an important role  in the enhancements induced by magnon nesting, unlike the contribution of $H^{xy}_4$, which is minor due to is particular dependence on magnon momenta.}
    
    {In connection to our results in the following sections, it is helpful to discuss the role of the spin size $S$ in the dynamics and to clarify the motivation for introducing the particular scaling of the coupling with $S$ in Eq.~\eqref{H_XXZ}. First, this scaling leads to a magnon dispersion that is independent of $S$, as seen in Eq.~\eqref{dispersion}. In particular, the magnon gap, which is located at $\bm{k}=0$ for $J>0$, does not depend on $S$. As a result, the location of the ground-state transition is independent of $S$ as long as fluctuations arising from interactions are neglected. This scaling therefore provides a natural way to compare our results across different system sizes.}

{Second, Eq.~\eqref{Jz_q} shows that magnon nonlinearities are weaker for larger spin sizes. A direct consequence is that magnon scattering processes are suppressed for larger $S$, which strengthens the correspondence between spins and nearly free magnons. A nonperturbative way to see this more clearly is to note that $\langle S^z_i \rangle \le S$, which implies, using Eq.~\eqref{HP},
\begin{equation}\label{eq:upper_bound}
    \frac{1}{N}\sum_{\bm{k}} \langle a^\dagger_{\bm{k}} a_{\bm{k}} \rangle \le S,
\end{equation}
and therefore places an upper bound on the total magnon population. As the number of magnons grows, for example due to external driving, magnon nonlinearities, including those in Eq.~\eqref{eq:H4} as well as higher-order terms omitted in the bosonic expansion of the spins, become increasingly important in order to enforce the bound in Eq.~\eqref{eq:upper_bound}.}
	
	\section{External drive}\label{sec:external_drive}
	We now proceed to describe the driving protocol and its effect on the dynamics of the system. In reality, an external driving field, such as a laser pulse, couples to various degrees of freedom in a material including charge, spin and vibrational lattice modes (phonons). We assume that the dominant effect on spins is captured by the parametric drive, which creates pairs of magnons at opposite momenta $\sim a^\dagger_{\boldsymbol{k}}a^\dagger_{-{\boldsymbol{k}}}$ and has been realized in numerous experiments~\cite{Demidov_magnonBEC2007,Serga_MagnonBEC2014,Zhou_magnon_scattering2021,Shan_parametric2024}. We show that the primary magnon instability leads to a secondary instability upon tuning the driving frequency.

	In the simplest case we can model this effect by a time-dependent Hamiltonian given by
	\begin{equation}\label{H_drive}
		H_\mathrm{drive}(t)= \frac{\xi}{4S}\cos(\omega t) \sum_i \Big[(S^+_i)^2+(S^-_i)^2\Big].
	\end{equation}
	$\xi$ is the amplitude of the drive, which is determined by the intensity of the driving field, $\omega$ is the driving frequency. In this work we use a semi-classical treatment assuming that $S\gtrsim 1$. We note that $(S^\pm)^2=0$ for $S=1/2$, so there would be no driving. In that case, parametric driving could be modeled by replacing the on-site operators in Eq.~\eqref{H_drive} with couplings of spins at different sites, i.e., $\sum_i (S^\pm_i)^2 \to \sum_{ij}t_{ij}S^+_i S^+_j$. As we will clarify later, such a modification does not qualitatively alter our results since they mostly rely on the energetics, rather than the geometric dependence of the drive. In Appendix~\ref{app:pulse}, we extend our analysis to include driving protocols involving localized pulses.

	\section{Primary instability: resonant magnon production}\label{sec:transverse}
	We characterize the effects of parametric driving on the system, with a particular focus on triggering dynamical instabilities. We consider the transverse two-point correlation function as given by
	\begin{equation}\label{C_trans}
		C^\perp_{\boldsymbol{k}} (t,t')= C^{xx}_{\boldsymbol{k}}(t,t') + C^{yy}_{\boldsymbol{k}}(t,t'),
	\end{equation}
    in terms of
	\begin{equation}
		C^{\alpha \alpha}_{\boldsymbol{k}}(t,t')=\frac12 \expval{\acomm{S^\alpha_{\boldsymbol{k}}(t)}{S^\alpha_{-\boldsymbol{k}}(t')}},
	\end{equation}
	where $\acomm{\cdot}{\cdot}$ is the anti-commutator and the Fourier transforms of spins is defined as $S^\alpha_{\boldsymbol{k}}=\sum_i e^{-i{\boldsymbol{k \cdot r_i}}} S^\alpha_i/\sqrt{N}$. According to Eq.~\eqref{HP}, the transverse spin correlation function is related to the magnon occupation by $C^\perp_{\boldsymbol{k}}(t,t)\approx  S\big(2 n_{\boldsymbol{k}}(t)+1\big)$. We consider the rate of change in the number of magnons $N=\sum_{\boldsymbol{k}} n_{\boldsymbol{k}}$ induced by the drive. {To the quadratic order in terms of the Holstein-Primakoff bosons, the drive Hamiltonian in Eq.~\eqref{H_drive} can be written 
    \begin{equation}
        H_\mathrm{drive}(t)\approx \frac12 \xi\cos(\omega t) \sum_{\bm{k}} \qty(a^\dagger_{\bm{k}}a^\dagger_{\bm{-k}} + a_{\bm{-k}}a_{\bm{k}} ).
    \end{equation}
    This shows the tendency of the drive to create and destroy pairs of magnons at opposite momenta.} Treating drive as a perturbation and starting from the ground state, we obtain the Fermi's golden rule
	\begin{equation}\label{Ndot_pairs}
		\dv{N}{t} \approx \frac{\pi\xi^2}{2} \mathcal{A}_\mathrm{pair}(\omega),
	\end{equation}
	where $\mathcal{A}_\mathrm{pair}$ is the two-magnon spectral density, which takes a simple form for nearly-free magnons:
	\begin{equation}\label{A_pair_free}
		\mathcal{A}_\mathrm{pair}(\omega)=\sum_{\boldsymbol{k}} \delta\big(\omega - \epsilon({\boldsymbol{k}}) - \epsilon(-{\boldsymbol{k}})\big).
	\end{equation}
	According to Eq.~\eqref{A_pair_free}, the drive generates pairs of magnons with zero center-of-mass momentum (Fig~\ref{fig:cartoon}b) whose energies satisfy
    \begin{equation}\label{eq:reson_cond}
        \omega =  \epsilon({\boldsymbol{k}}) + \epsilon(-{\boldsymbol{k}}),
    \end{equation}
    and the absorption rate is proportional to the available phase space to create such magnon pairs. In the case of a generalized quadratic drive given by $H_\mathrm{drive}\sim \sum_{ij} t_{ij}S^+_i S^-_j$, the momentum summation in Eq.~\eqref{A_pair_free} will be modified to $
	\mathcal{A}_\mathrm{pair}(\omega)\propto\sum_{\boldsymbol{k}} |t_{\boldsymbol{k}}|^2 \,\delta\big(\omega - \epsilon({\boldsymbol{k}}) - \epsilon(-{\boldsymbol{k}})\big)$,
	which does not alter our results qualitatively, as they mainly rely on the energy condition imposed by the Dirac delta. Coherent magnon production by parametric driving has been experimentally realized in various magnonic systems~\cite{Demidov_magnonBEC2007,Serga_MagnonBEC2014,Zhou_magnon_scattering2021,Shan_parametric2024}.
	Eq.~\eqref{Ndot_pairs} is also commonly interpreted as a parametric instability of magnons as classical spin waves~\cite{Stancil_2009spinwaves,lvov_2012wave}, in which the amplitude of a wave grows exponentially under a parametric drive whose frequency matches twice the natural frequency of the mode.  At later times, magnon nonlinearities  become important and the free magnon approximation becomes inaccurate. This requires a non-perturbative approach to dynamics whose numerical results will be shown later.

	\begin{figure*}[!t]
		\centering
		\includegraphics[height=.22\linewidth]{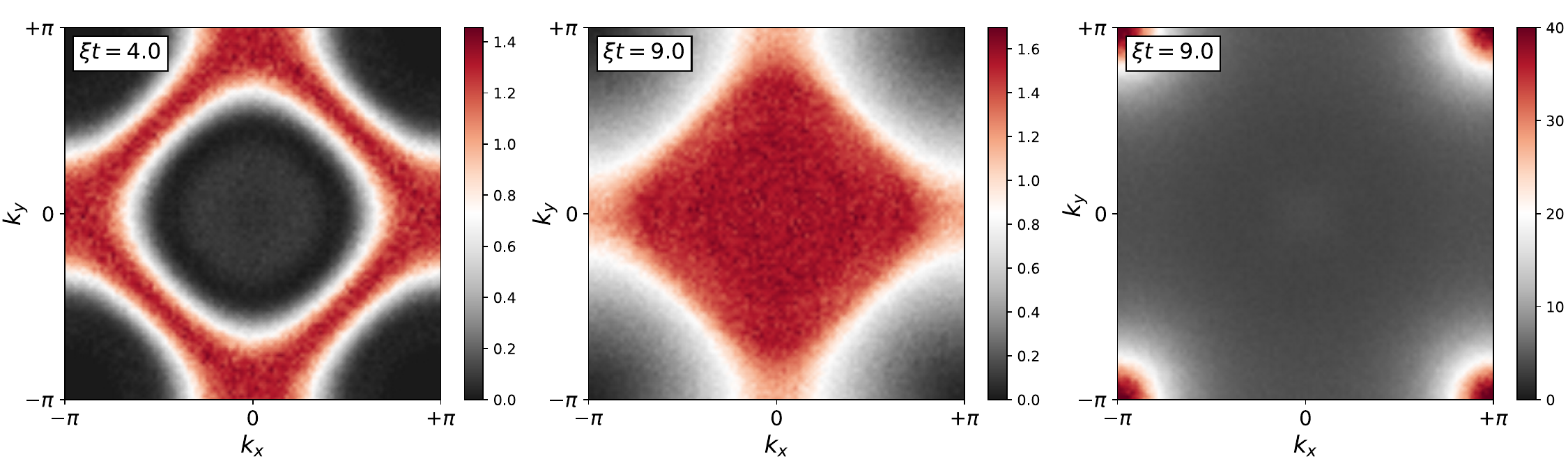}

		\caption{\textbf{Magnon nesting for AFM coupling.} (Left) Early-time magnon occupation in the momentum space due to parametric driving at the nesting frequency. (Center)  Redistributed magnons at later times due to scattering processes, resembling a Fermi surface. (Right) Longitudinal correlation function ($C^{zz}_{\boldsymbol{k}}(t,t)$), showing AFM instability at $(\pi,\pi)$. The system consists of $100\times100$ spins, and the other parameters are  $J/h=0.2$, $\Delta=-2$, $\xi/h=0.1$ and $S=5$.}
		\label{fig:xx_zz_AFM}
	\end{figure*}

	\section{Secondary instability due to magnon nesting}\label{sec:longitudinal}
	Longitudinal modes correspond to the component of spin parallel to the direction of magnetization, which in our case is $S^z$ (Eq.~\eqref{psi0}). These can be characterized by the following correlation function
	\begin{equation}\label{C_zz_def}
		C^{zz}_{\boldsymbol{q}}(t,t')=\frac12 \expval{\acomm{S^z_{\boldsymbol{q}}(t)}{S^z_{-{\boldsymbol{q}}}(t')}}_c,
	\end{equation}
	 which can be expressed in terms of the magnon density operator as
    $C^{zz}_{\boldsymbol{q}}(t,t')= \langle\{\rho_{\boldsymbol{q}}(t),\rho_{-{\boldsymbol{q}}}(t')\}\rangle/2$.
    As a first approximation, we apply the Wick's decomposition to the above correlation function to get (for $t=t'$) 
    \begin{equation}\label{C_zz_free}
        \tilde{C}^{zz}_{\boldsymbol{q}}(t,t)=\frac{1}{N}\sum_{\boldsymbol{k}} \Big( n_{\boldsymbol{k+q}}(t)\,n_{\boldsymbol{k}}(t)+ n_{\boldsymbol{k}}(t)  \Big).
    \end{equation}
    We call this approximation the truncated Gaussian approximation, as we only take into account the early-time profile of magnon population, rather than employing a systematic variational Gaussian wavefunction for bosons as discussed, for instance, in Ref.~\cite{Guaita_Gaussian2019}. 
    The first term on the right-hand side of Eq.~\eqref{C_zz_free} provides an enhanced contribution when a single momentum $\boldsymbol{q}$ connects a large number of highly populated momentum pairs. This state can be created by external driving when the resonance condition in Eq.~\eqref{eq:reson_cond} is simultaneously satisfied. For the square lattice, the nesting vectors are   well known to be~\cite{fradkin2013field,Sachdev_2011}
    \begin{equation}
        \boldsymbol{q}=(\pi,\pm\pi),
    \end{equation}
    which inter-connect the nested momenta satisfying
	\begin{equation}\label{nested_K}
		 \abs{k_x\pm k_y}\approx\pi.
	\end{equation}
	This corresponds to the edges of a square in the Brillouin zone whose corners are located at $(\pm \pi,0)$ and $(0,\pm \pi)$, also shown in the middle of Fig.~\ref{fig:cartoon}c. To populate  the nested modes, the driving frequency should be close to the nesting frequency
    \begin{equation}\label{omega_nest}
        \omega_\mathrm{nest}=2\epsilon(\boldsymbol{k})|_{|k_x \pm k_y|=\pi}=2(h+\Delta J).
    \end{equation}
    Intuitively, the drive creates a squeezed state of bosons at $(\boldsymbol{k},-\boldsymbol{k})$ on the nested contour with a momentum difference of $\boldsymbol{q}=(\pi,\pm\pi)$. A phase coherence locking between these two particles favors the formation of a ‘condensate’ of a bi-linear of bosons, $\langle\rho_{\boldsymbol{q}}\rangle$, which is therefore modulated with a spatial period of $\sim 1/\boldsymbol{q}$. This explains the early-times peak in $\tilde{C}^{zz}_{\boldsymbol{q}=(\pi,\pi)}$ (Appendix~\ref{app:TGA}), which can be thought of as developing coherence between bosonic states that differ by $(\pi,\pi)$. This mechanism is in strong resemblance with pattern formation in two components Bose mixtures of ultra-cold atoms, see for instance~\cite{sowinski2019one}.

    Naturally, the approximation used above has its own limitations. It is insensitive to the value of $\Delta$, and, due to the $1/N$ prefactor in Eq.~\eqref{C_zz_free}, the enhancement from nesting results only in a small correction to longitudinal correlations. Furthermore, it always has a peak at $\boldsymbol{q}=0$, which is actually stronger than its value at $\boldsymbol{q}=(\pi,\pi)$, even when magnons are nested.  
    The observed enhancement exhibits a strong dependence on $\Delta$ and becomes suppressed for $|\Delta| \lesssim 1$. Furthermore, the numerical values of longitudinal correlations (next section) significantly exceed the truncated Gaussian prediction. All these features are documented in detail in Appendix~\ref{app:TGA}.

    \section{Bose-Stoner dynamical instability}\label{sec:Bose_stoner}
    A thorough description of the dynamics requires incorporating the effect of interactions. {At the leading order in the number of magnons, interactions are given by Eq.~\eqref{eq:H4}. The two terms in $H_4$ contribute differently to the instability at $\bm{q}=(\pi,\pi)$. As we discussed previousuly, the instability originates from magnons at momenta near nesting, which satisfy Eq.~\eqref{nested_K}. However, since $H_4^{xy}$ in Eq.~\eqref{eq:H4_xy} is particularly suppressed for $\bm{k,p}$ near nesting, it has a weaker effect compared to $H_4^{z}$ (Eq.~\eqref{eq:H4_z}), which depends on the exchanged momentum $\bm{q}$. Consequently, we ignore $H_4^{xy}$ in our discussion below, which also simplifies our treatment. Our numerical results in Section~\ref{sec:results} do not rely on this assumption and show the qualitative validity of our arguments.}
    
    As shown in Appendix~\ref{app:RPA}, applying the random phase approximation (RPA)~\cite{altland2010condensed} yields the following expression for the correlation function:
    \begin{equation}\label{C_zz_RPA_t}
        C^{zz}_{\boldsymbol{q}}(t,t) = \int_{-\infty}^{t} \int_{-\infty}^t \Gamma_{\boldsymbol{q}}(t,\tau_1) \,\Gamma^*_{\boldsymbol{q}}(t,\tau_2)\,\tilde{C}^{zz}_{\boldsymbol{q}}(\tau_1,\tau_2) \,d\tau_1 d\tau_2 ,
    \end{equation}
    where
    \begin{equation}
        \Gamma^{-1}_{\boldsymbol{q}}(t,t') = \delta(t-t') + \Delta J(\boldsymbol{q}) \Pi_{\boldsymbol{q}}(t,t'),
    \end{equation}
      is a dynamical enhancement kernel in terms of $J(\bm{q})$ in Eq.~\eqref{Jz_q} and the magnon polarization function $\Pi_{\boldsymbol{q}}$ is given by (Appendix~\ref{app:RPA})
    \begin{equation}\label{pi_free}
		\Pi_{\boldsymbol{q}}(\omega)=-\frac{1}{N} \sum_{\boldsymbol{k}} \frac{
			n_{\boldsymbol{k+q}} - n_{\boldsymbol{k}}
		}{\omega -\epsilon({\boldsymbol{k+q}}) + \epsilon({\boldsymbol{k}}) + i 0^+}.
	\end{equation}
      According to Eq.~\eqref{C_zz_RPA_t}, the truncated Gaussian approximation remains valid in the weak-coupling limit $\Delta \ll 1$, where $\Gamma_{\boldsymbol{q}}(\omega)\approx \delta(t-t')$. However, under nesting conditions $\Gamma_{\boldsymbol{q}}(\omega)$ is strongly enhanced for $\boldsymbol{q}\approx (\pi,\pi)$ and at low frequencies. This behavior can also be understood from the frequency-domain representation of Eq.~\eqref{C_zz_RPA_t}, yielding
    \begin{equation}\label{C_zz_RPA_w}
    C^{zz}_{\boldsymbol{q}}(\omega)=\frac{\tilde{C}^{zz}_{\boldsymbol{q}}(\omega)}{|1+\Delta J(\boldsymbol{q})\Pi_{\boldsymbol{q}}(\omega)|^2}.
    \end{equation}
    The correlation function signals an instability at a wave vector ${\boldsymbol{q}}$, provided that
	\begin{equation}\label{rpa_condition}
		1+ \Delta J({\boldsymbol{q}})\Pi_{\boldsymbol{q}}(\omega)|_{\omega\to0}=0,
	\end{equation}
	 which is the generalized Stoner criterion for finite wave-vectors~\cite{kubler2021theory,Sachdev_2023}. {Upon increasing the coupling $J$ from small values, the instability is expected to emerge at wave vectors that satisfy Eq.~\eqref{rpa_condition} first, which depends on $\bm{q}$ through both $J(\bm{q})$ and $\Pi_{\bm{q}}$. In general, modes with larger $|J(\bm{q})|$ are expected to be more susceptible to the instability. The symmetries of the system can also play an important role. For example, if a particular wave vector $\bm{q}$ is the most enhanced mode, then all wave vectors related to $\bm{q}$ by the point-group symmetries of the lattice are equally likely to develop an instability. This produces a competition among symmetry-related modes which, in thermal equilibrium in fermionic systems, typically leads to spontaneous symmetry breaking that selects one of these momenta as the ordering wave vector of a charge-density wave. Far from equilibrium, we expect this competition to render the enhancement more fragile and less pronounced, or even fully suppress it.
}

     {When the system is driven close to perfect nesting, that is, when the momenta near Eq.~\eqref{nested_K} are excited, the wave vector $\boldsymbol{q} = (\pi,\pi)$ is the most likely to exhibit the instability based on the arguments given above. First, the magnitude of $J(\boldsymbol{q})$ is maximized at this wave vector. Second, $\boldsymbol{q} = (\pi,\pi)$ is mapped onto itself by all symmetry operations of the square lattice, which eliminates competition among symmetry-related modes that could otherwise weaken the instability. For these reasons, we focus on this wave vector in the following. Our numerical results in Section~\ref{sec:results} confirm that the enhancement is strongest for momenta in its vicinity.}
     
     For $\boldsymbol{q} = (\pi, \pi)$, the interaction satisfies $J(\boldsymbol{q}) < 0$, so the sign of $\Delta \,\Pi_{\boldsymbol{q}}(0)$ determines whether correlations are enhanced by reducing the denominator in Eq.~\eqref{C_zz_RPA_w}. For nested momenta satisfying Eq.~\eqref{nested_K}, the denominator becomes small when $\omega \to 0$, leading to stronger interaction effects, similar to the instabilities of nested Fermi surfaces~\cite{auerbach2012interacting,Sachdev_2023}. In the time domain, the this argument translates to a slow decay of $\Gamma_{\boldsymbol{q}}(t,t')$, reflecting significant temporal memory. As a result, the time integrals in Eq.~\eqref{C_zz_RPA_t} accumulate correlations over extended durations, leading to a pronounced dynamical enhancement of $C^{zz}_{\boldsymbol{q}}(t,t)$.

    \begin{figure*}[!t]
		\centering
    \includegraphics[height=.22\linewidth]{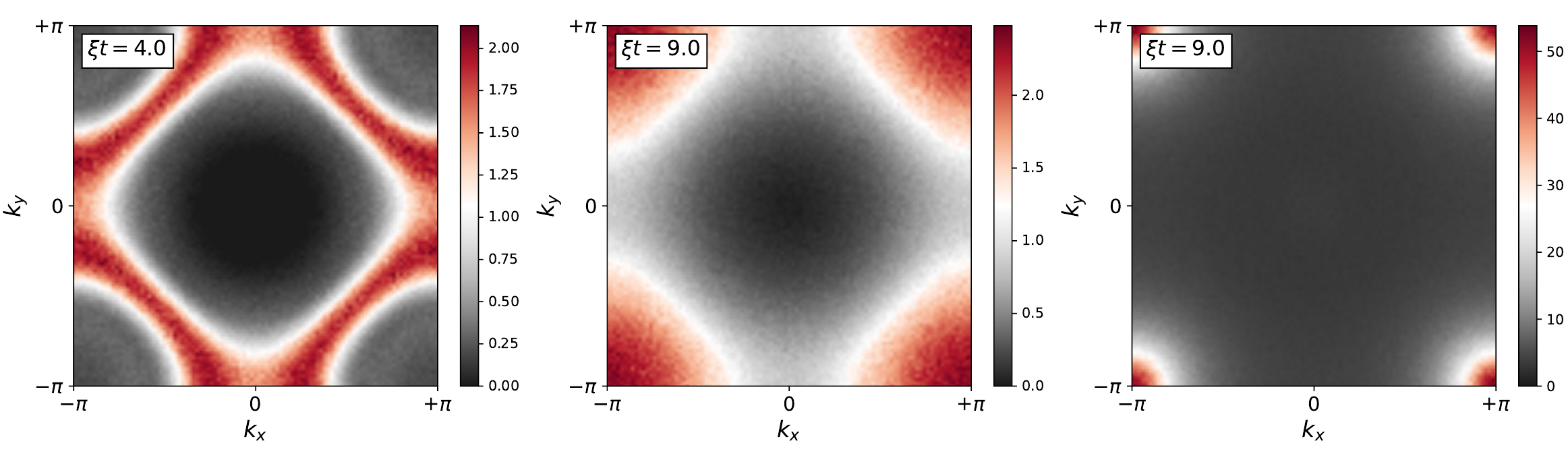}
		\caption{\textbf{Magnon nesting for FM coupling.} (Left) Early-time magnon occupation in the momentum space due to parametric driving at the nesting frequency. (Center)  Magnon scattering modifies the distribution, which now resembles an inverted  Fermi surface. (Right) Longitudinal correlation function ($C^{zz}_{\boldsymbol{k}}(t,t)$), showing AFM instability at $(\pi,\pi)$. Except for $\Delta=2$, the other parameters are the same as in Fig.~\ref{fig:xx_zz_AFM}.}
		\label{fig:xx_zz_FM}
	\end{figure*}
    
    When $\Delta < 0$, corresponding to interactions that favor antiferromagnetic (AFM) order, enhancement occurs if $\Pi_{\boldsymbol{q}}(0) < 0$. According to Eq.~\eqref{pi_free}, this condition is met when the magnon distribution is such that lower-energy nested modes are more populated than higher-energy ones. In this case, the momentum distribution of magnons resembles that of fermions near a nested Fermi surface. For $\Delta > 0$, corresponding to interactions that favor ferromagnetic (FM) correlations, enhancement instead requires $\Pi_{\boldsymbol{q}}(0) > 0$. This condition is satisfied when the magnon population is inverted compared to the AFM case, with higher-energy modes more occupied than lower-energy ones, analogous to an inverted Fermi surface. As we show numerically below, both types of distributions naturally emerge through magnon scattering. In particular, the AFM-enhancing distribution arises even when the underlying interaction is ferromagnetic ($\Delta > 0$), indicating that the secondary instability is not constrained by the sign of the equilibrium coupling.

      We now explain the significance of enhanced correlations at ${\boldsymbol{q}}=(\pi,\pi)$. Fourier transforming back to the real space, it indicates the emergence of a spin pattern with a periodicity of two sites, i.e., alternating magnetization between the odd and even sub-lattices of the square lattice. Moreover, it can be regarded as a secondary instability, which emerges on top of the primary parametric instability which is directly triggered by the drive. The role of the primary instability is to create a non-equilibrium magnon state which is macroscopically degenerate with respect to magnon scatterings that satisfy Eq.~\eqref{nested_K}. This highly degenerate state is then prone to lift its degeneracy by developing the secondary AFM instability. While the primary instability can be fully explained using a single-particle picture, the latter emerges from non-equilibrium many-body correlations, with no equilibrium bosonic counterparts.

	\section{Real-time dynamics}\label{sec:results}
	So far, we have analytically illustrated the possibility of creating secondary instabilities in a parametrically driven XXZ model. It is necessary to obtain the real-time dynamics of the system in order to evaluate the predicted dynamical instabilities. In this work, we employ the truncated Wigner approximation (TWA), derived from a phase-space representation of quantum mechanics~\cite{POLKOVNIKOV2010,Schachenmayer_dtwa,Sundar_twa,Huber_DDTWA,Fleischhauer_2022,hosseinabadi2025_TWA,rodriguez2022far}, which offers the dual advantages of being non-perturbative and scalable to large system sizes.

    For $\Delta<0$, corresponding to an AFM coupling, the distributions of magnons at early times is shown in the left panel of Fig.~\ref{fig:xx_zz_AFM}, where $n_{\boldsymbol{k}}$ grows according to the Fermi's golden rule, and magnons proliferate at pairs of opposite momenta which satisfy Eq.~\eqref{nested_K}. This is the primary instability which was discussed earlier. At later times, as the number of pumped excitations grows significantly, magnon nonlinearities become important and begin to reshape the distribution of excitations. In particular, we observe that a large fraction of magnons are scattered into the interior of the diamond-shaped contour, as shown in the middle panel of Fig.~\ref{fig:xx_zz_AFM}, and $n_{\boldsymbol{k}}$ transiently resembles the nested distribution characteristic of fermions on a half-filled square lattice. This configuration provides the appropriate excitation profile to trigger the secondary Bose-Stoner instability, which can be seen from the sharp peak of longitudinal correlations at $\boldsymbol{q}=(\pi,\pi)$, as shown in the right panel of Fig.~\ref{fig:xx_zz_AFM}.

 \begin{figure*}[!t]
    \centering
    \includegraphics[width=.98\linewidth]{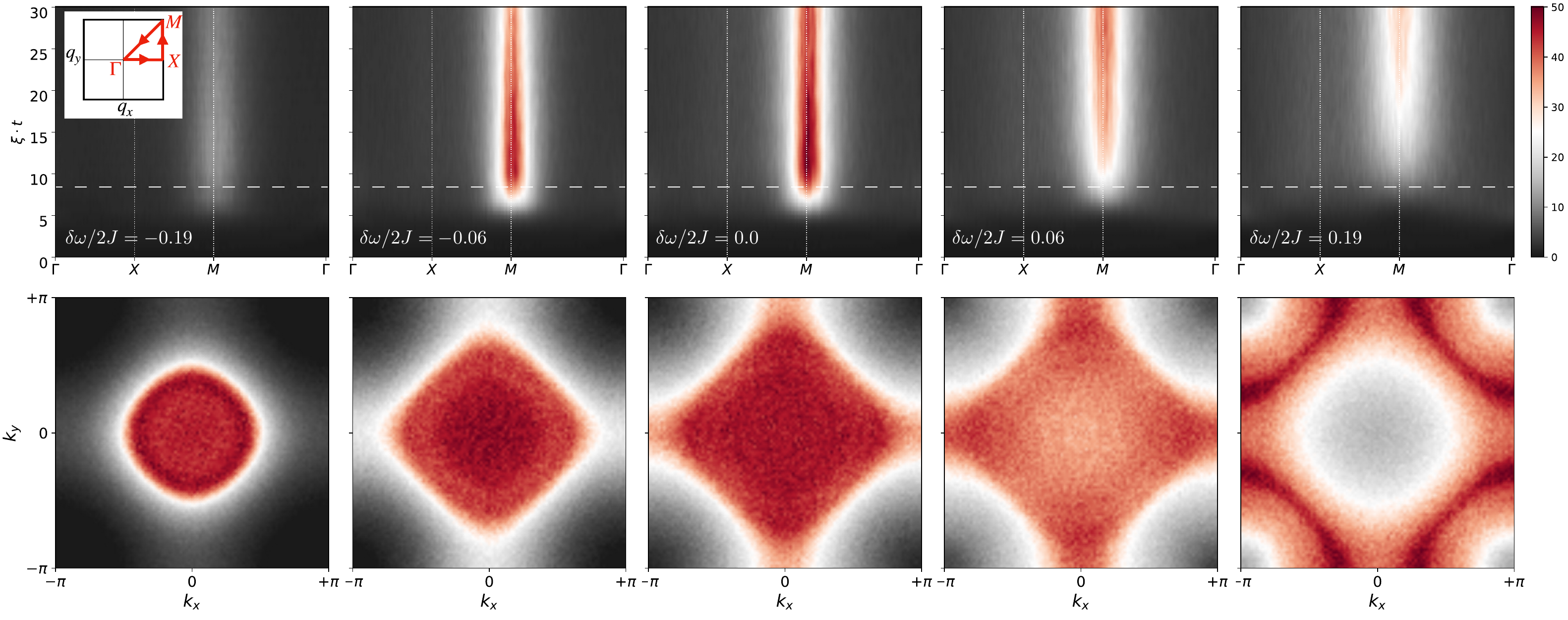}
    \caption{ (Top row) The time evolution of $C^{zz}_{\boldsymbol{q}}(t,t)$ for various driving frequencies (measured relative to the nesting frequency) is shown for momenta along a closed path through the Brillouin zone (depicted in the inset). Close to the nesting frequency, a secondary instability is triggered. (Bottom row) Magnon population at $\xi t=9.0$ (marked by dashes in the top row) for driving frequencies given in the top row, demonstrating the formation of instability close to nesting. The color-scales in the bottom panel are different, but have the same order of magnitude with $1.5 \lesssim n_\mathrm{max} \lesssim 3$ across different figures. The other parameters are the same as in Fig.~\ref{fig:xx_zz_AFM}.}
    \label{fig:c_contour}
	\end{figure*}

    We emphasize that, the filling of the nested contour's interior arises specifically due to the finite value of $\Delta$. For $\Delta = 0$, corresponding to the XX Hamiltonian in a transverse field (cf. Eq.~\eqref{H_XXZ} with $\Delta=0$), the magnon populations resulting from two drives with opposite detunings from the nesting frequency, $\omega_\pm = \omega_\mathrm{nest} \pm \delta\omega$, are related by a momentum shift of $\boldsymbol{q} = (\pi,\pi)$. Thus, driving the system exactly at the nesting frequency ($\delta\omega = 0$) should not inherently favor the interior over the exterior regions of the nesting diamond. As shown in Appendix~\ref{app:chiral}, this behavior of the XX Hamiltonian can be understood via a ``chiral" transformation given by the unitary operator
    \begin{equation}\label{eq:chiral_trans}
        U = \exp\left[-\frac{i\pi}{2} \sum_{j_x, j_y} (-1)^{j_x + j_y} S^z_j \right],
    \end{equation}
    where $j_x$ and $j_y$ are the lattice coordinates. This operator applies a $\pi/2$ rotation in opposite directions to spins on even and odd sublattices. This operator reverses the sign of $\delta\omega$ and shifts the momentum of spin-wave operators as
    \begin{equation}\label{eq:chiral_S}
        U S^+_{\boldsymbol{k}} U^\dagger = -i \, S^+_{\boldsymbol{k} + (\pi, \pi)}.
    \end{equation}
    For $\Delta \neq 0$, this chiral mapping between opposite values of $\delta\omega$ is broken. As a result, the interior and exterior of the nesting contour are no longer equivalent when $\omega = \omega_\mathrm{nest}$.

   We now turn to the case of ferromagnetic coupling ($\Delta > 0$). As in the antiferromagnetic case, the early-time magnon distribution is governed by the primary parametric instability, as shown in the left panel of Fig.~\ref{fig:xx_zz_FM}. At later times, however, scattering processes redistribute the magnons differently than in the $\Delta < 0$ case discussed above. Specifically, as shown in the middle panel of Fig.~\ref{fig:xx_zz_FM}, magnons predominantly scatter into higher-energy modes, resulting in a distribution that resembles an inverted nested Fermi surface. Under these conditions, the polarization function $\Pi_{\boldsymbol{q}}(0)$ in Eq.~\eqref{pi_free} becomes positive, enabling the enhancement of correlations, \emph{despite the fact that the microscopic couplings favor purely ferromagnetic order}. This behavior is confirmed by the numerical growth of longitudinal correlations, shown in the right panel of Fig.~\ref{fig:xx_zz_FM}.

	To demonstrate that the peak is genuinely a feature of nested magnons, we have shown the time-evolution of $C^{zz}$ for different values of $\delta\omega = \omega - \omega_\mathrm{nest}$ along a closed path through the Brillouin zone, along with a snapshot of magnon distribution for each case, in the top and bottom rows of Fig.~\ref{fig:c_contour}, respectively. We see the peak at ${\boldsymbol{q}}=(\pi,\pi)$ disappears for drives far away from nesting. While these results are for $\Delta<0$, we observed similar behavior for $\Delta>0$. The instability appears across a finite frequency domain around $\omega_\mathrm{nest}$, and there is no need to fine-tune the driving frequency. This can be seen also from the dependence of $C^{zz}_{(\pi,\pi)}(t,t)$ on the driving frequency, as shown in Fig.~\ref{fig:cartoon}c.

    We refer the reader to Appendix~\ref{app:heat} for a detailed discussion of the system’s long-time dynamics and heating under continuous driving, as well as an extension of our results to temporally localized pulse protocols.

	\section{Experimental Realizations and generality of the mechanism}\label{sec:experiment}

    In the following, we provide an overview of experimental systems that could potentially host the non-equilibrium boson nesting and its resulting instabilities discussed in this work. A key requirement for any relevant physical system is the presence of effective bosonic degrees of freedom on a lattice whose geometry supports nested regions in the momentum space. Below, we discuss several such platforms in more detail.

	(i) Solid state magnetic materials are the natural platforms for studying spin models. Particularly relevant to our work are monolayer or few-layer 2D or quasi-2D magnets, such as van der Waals materials~\cite{Vertikasari_2Dmagnets}. Another example includes conventional magnetic insulators, such as yttrium iron garnet (YIG). The primary magnon instability can be efficiently induced in these systems through external driving~\cite{Demidov_magnonBEC2007,Serga_MagnonBEC2014,Zhou_magnon_scattering2021,Shan_parametric2024}, and by tuning the driving frequency, one can produce a nested magnon distribution. We remark that, the geometry of the lattice determines the shape of this distribution, which may differ from the diamond-like structure observed in square lattices. Consequently, the secondary instability may manifest at momenta different from $\boldsymbol{q}=(\pi,\pi)$.

	(ii) Quantum simulators  provide a unique opportunity to engineer many-body systems that closely follow the physics of the target model Hamiltonian. Particularly relevant to our work is the Google Quantum AI platform introduced in Ref.~\cite{GoogleQuAI_2024}, which successfully simulates the dynamics of a two-dimensional quantum XY model. These platforms allow for probing the theoretical predictions of this work by engineering the easy-axis $S^z_i S^z_j$ coupling of Eq.~\eqref{H_XXZ}. An advantage of quantum simulators compared to magnetic materials is the absence of phonons, which can significantly alter the dynamics in magnetic systems, especially at later times. In magnetic materials, spin-phonon coupling introduces unwanted heating, complicating the study of intrinsic spin interactions. Quantum simulators circumvent this issue, offering a cleaner and more controlled environment to investigate quantum many-body phenomena.
    
	(iii) Ultracold bosons on optical lattices. While we discussed a spin system in this work, one can also consider the phenomenon of boson nesting as a generalization of magnon nesting. According to the picture provided in this work, the presence of a drive term creating pairs is necessary to trigger the nesting instability.  {The pair creation can be achieved, for instance, by shaking optical lattices loaded with Bose-Einstein-condensates, leading to the temporal modulation of the tunneling amplitude. As was theoretically proposed in Refs.~\cite{Kramer_parametricBEC2005,Tozzo_parametricBEC2005,bukov_drivenBEC2015,Lellouch_parametric2017} and experimentally verified in Refs.~\cite{Boulier_parametricHeating2019,Wintersperger_parametric2020}, pairs of 
     Bogoliubov quasi-particles with opposite momenta are created in these systems upon tuning the driving frequency, which parallel our findings about magnons. Building on this, our work suggests the emergence of nesting-induced dynamical enhancement of density correlations in driven superfluids on optical lattices, provided that the system has interactions which favor density-wave ordering. The driven nested state, enables these interactions to dominate and trigger a secondary instability in the system.

	\section{Discussion}\label{sec:conclusion}
	
	We studied the non-equilibrium dynamics of a spin system on a two-dimensional square lattice subject to a harmonic parametric drive. Combining physical arguments with numerical simulations, we predicted the emergence of two distinct dynamical instabilities induced by the drive. We showed that a primary instability, characterized by the rapid proliferation of magnon pairs with opposite momenta, is triggered when the driving frequency matches the energy required to create a magnon pair~\cite{Demidov_magnonBEC2007,Serga_MagnonBEC2014,Zhou_magnon_scattering2021,Shan_parametric2024}. This results in the accumulation of magnons at specific wave vectors where their energies equal half the driving frequency, forming bright contours in momentum space. At later times, magnon nonlinearities become significant and give rise to scattering processes that redistribute the magnon population across momentum space.

	Next, we predicted the emergence of a secondary instability when the growing magnon modes from the primary instability form a nested contour in momentum space. Nesting occurs when a macroscopically large number of amplified momenta are connected by the same wave vector, analogous to the nesting condition in two-dimensional fermionic systems. Our numerical simulations, based on the truncated Wigner approximation, confirmed this prediction and revealed a strong enhancement of antiferromagnetic correlations when the driving frequency is tuned near the nesting value. { The secondary AFM instability originates from magnon nonlinearities, and \emph{emerges for both FM and AFM couplings in the Hamiltonian}.}     {Remarkably, similar behavior has been observed in transport experiments on ultracold two-dimensional fermions, where the dynamics appear insensitive to the sign of interactions as a result of a hidden symmetry~\cite{schneider2012fermionic}. This parallel suggests that experiments in atomic, molecular and optical physics could soon be used to investigate the phenomena discussed here, extending the relevant platforms for our results  beyond quantum materials and out-of-equilibrium solid-state systems.  }

	The findings of this paper open several promising directions for future investigation,  aimed at testing the robustness and broader applicability of our results across various state-of-the-art platforms. A natural avenue is to examine how non-equilibrium magnon nesting is modified when  phonons or spin-orbit coupling are included, thereby moving closer to realistic material implementations.
Another important question is the role of short-range interactions in our model. For instance, replacing the uniform planar couplings in Eq.~\eqref{H_XXZ} with interactions that decay as a power law with distance~\cite{defenu} could shed light on how locality influences magnon nesting.
In a similar spirit, it would be valuable to explore whether analogous nesting phenomena can be realized in parametrically driven Bose-Hubbard models. This would help clarify the importance of spin nonlinearities in supporting nesting, in contrast to the quartic nonlinearities characteristic of Hubbard-type hamiltonians. Its outreach to quantum simulators based on superconducting qubits would be natural.
 On the theoretical side, an intriguing question is how essential coherent driving is for realizing magnon nesting. Could a similar effect be achieved using an incoherent magnon pump? Exploring this possibility would offer insights into how correlations can emerge in many-body systems without any a priori imposed coherence.

	\section{ACKNOWLEDGEMENTS}
	The authors thank M. Buchhold,   M. Hafezi, M. Mitrano and R. Moessner for useful discussions. H.H. and J.M. were financially supported by the Deutsche Forschungsgemeinschaft (DFG, German Research Foundation) through TRR 288 - 422213477 (projects B09A) and the Dynamics and Topology Centre, funded by the State of Rhineland Palatinate.  Y.T. is supported by NSF under Grant No. DMR-2049979. E.D is supported by the SNSF project 200021 212899, the Swiss State Secretariat for Education, Research and Innovation (SERI) under contract number UeM019-1 and NCCR SPIN, a National Centre of Competence in Research, funded by the Swiss National Science Foundation (grant number 225153). H.H. gratefully acknowledges the computing time granted on the supercomputer MOGON 2 at Johannes Gutenberg-University Mainz (hpc.uni-mainz.de). J.M. acknowledges the Pauli Center at ETH-Z for hospitality.

    \appendix

    \section{Pulse protocol}\label{app:pulse}

    In connection with the typical protocol of pump-probe experiments~\cite{delatorre_2021colloquium,tomeljak2009dynamics,fausti2011light,hu2014optically,Mitrano_lightinduced2016}, we consider a Gaussian pulse, whose spectrum is centered around $\omega_0$ and has a width $\sigma_\omega$, as given by
	\begin{equation}
		f(t)=\int_{-\infty}^{+\infty} f(\omega) \cos\big[\omega (t-t_0)\big]\,d\omega,
	\end{equation}
	where
	\begin{equation}\label{f_pulse}
		f(\omega) = \frac{1}{\sqrt{2\pi }\sigma_\omega} e^{-\frac{(\omega-\omega_0)^2}{2\sigma_\omega^2}}.
	\end{equation}
	 The frequency broadening $\sigma_\omega$ corresponds to a pulse with a temporal width of $\tau_\mathrm{pulse}\propto \sigma_\omega^{-1}$, such that the harmonic limit is recovered for $\tau_\mathrm{pulse}\to\infty$.  The delay parameter $t_0$ is chosen to satisfy $t_0 \gg \tau_\mathrm{pulse}$, such that the pulse  is applied after $t=0$.

     In this case, instead of the magnon creation rate in Eq.~\eqref{Ndot_pairs}, we can obtain the total number magnons created during the pulse from
	\begin{equation}\label{Ndot_pulse}
		\Delta N \approx \frac{\pi\xi^2}{4}  \int_{-\infty}^{+\infty} \abs{f(\omega)}^2 \mathcal{A}_\mathrm{pair}(\omega) \, d\omega.
	\end{equation}
	According to this expression, the energies and momenta of magnon pairs are determined according to the power-spectrum of the pulse $|f(\omega)|^2$.

    Now, we discuss our numerical results for the pulse protocol. According to Eq.~\eqref{Ndot_pulse}, we expect to get magnon nesting as long as the frequency spectrum of the pulse is sufficiently localized close to $\omega_\mathrm{nest}$. A broad-frequency pulse produces magnons across a wide range of energies and momenta, away from the nesting regime. Subsequently, these magnons can enhance unwanted fluctuations that compete with the AFM instability.

    \begin{figure}[!t]
		\centering
		\includegraphics[width=0.9\linewidth]{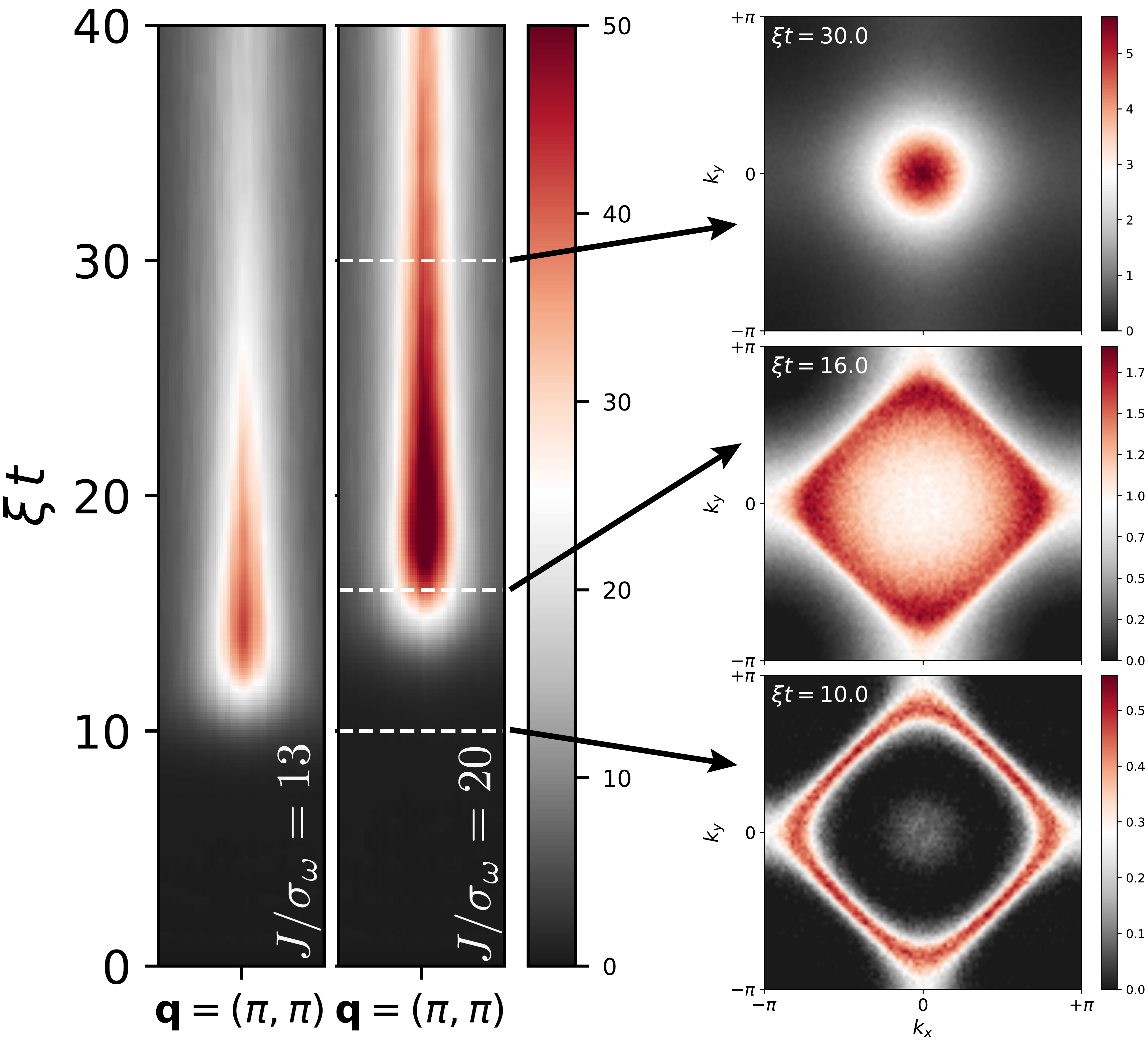}
		\caption{(Left panel) Profile of $C^{zz}$ close to ${\boldsymbol{q}}=(\pi,\pi)$ for two different pulses with the same mean-frequency $(\omega_\mathrm{nest}-\omega_0)/2J=0.1$ and different widths. Pulses that are broader in the time domain create stronger instabilities. (Right panel) Snapshots of magnon distribution at three different times which are marked by dashes in the left panel. Nesting is strongest near the onset of secondary instability. Other parameters are the same as in Fig.~\ref{fig:xx_zz_AFM}.}
		\label{fig:C_zz_pulse}
	\end{figure}

	In our numerical calculations, we consider pulses whose centers are close to $\omega_\mathrm{nest}$, and focus on the effect of frequency broadening $\sigma_\omega$ on the secondary instability. As shown in Fig.~\ref{fig:C_zz_pulse} (left panel), the nesting enhancement still occurs in this case. We see that narrow-frequency pulses yield stronger enhancement, as they excite the nesting wave-vectors with higher accuracy. The peak enhancement occurs earlier for temporally narrower pulses, simply because the pulse reaches its maximum amplitude more rapidly. In the right panel of Fig.~\ref{fig:C_zz_pulse}, we show the magnon occupation at early (bottom panel), intermediate (middle panel), and late (top panel) times. During the early-time dynamics, the system is dominated by the parametric generation of magnons. At intermediate times, magnon scattering leads to the formation of a nested distribution, which triggers the onset of the secondary instability. Finally, at late times, the majority of magnons are scattered toward small momenta, resulting in the suppression of the secondary instability.

	To summarize, our findings about the pulse protocol indicate that realizing the nesting instability is not sensitive to the details of the driving profile, as long as the nesting momenta are driven sufficiently stronger than the rest of of the modes. 

    \section{Longitudinal correlations in the truncated Gaussian approximation}\label{app:TGA}

We show that the truncated Gaussian approximation for $C^{zz}$, given in Eq.~\eqref{C_zz_free} and repeated here for convenience,
\begin{equation}\label{C_zz_free_app}
\tilde{C}^{zz}_{\boldsymbol{q}}(t,t) = \frac{1}{N} \sum_{\boldsymbol{k}} \left( n_{\boldsymbol{k+q}}(t) n_{\boldsymbol{k}}(t) + n_{\boldsymbol{k}}(t) \right),
\end{equation}
fails to capture the numerically observed dynamics. This discrepancy arises from the omission of quantum fluctuations inducing an enhancement of  response, as accounted for in Eqs.~\eqref{C_zz_RPA_t}~and~\eqref{C_zz_RPA_w}. To evaluate the right-hand side of Eq.\eqref{C_zz_free_app}, we extract magnon occupations from TWA simulations by computing transverse spin correlations and relating them to $n_{\boldsymbol{k}}$ via $C^\perp_{\boldsymbol{k}}(t,t)\approx  S\big(2 n_{\boldsymbol{k}}(t)+1\big)$. Substituting these occupations into the expression for $\tilde{C}^{zz}_{\boldsymbol{q}}(t,t)$ yields the truncated Gaussian approximation. 

For $\Delta=2$, the results are shown in the top row of Fig.~\ref{fig:gauss_vs_twa}. As argued in the main text, truncated Gaussian approximation consistently predicts a dominant peak at $\boldsymbol{q} = 0$. At early times, some enhancement appears near $\boldsymbol{q} = (\pi, \pi)$, which originates from momentum pairs $\boldsymbol{k}$ and $\boldsymbol{k} + \boldsymbol{q}$ in Eq.~\eqref{C_zz_free_app} that lie on opposite, parallel edges of the nesting square. Additional enhancement is also visible along the Brillouin zone diagonals, arising from pairs located on the same edge of the nesting contour. However, at later times, the $\boldsymbol{q} = 0$ component dominates, and no significant feature remains at $(\pi, \pi)$. These trends contrast sharply with the TWA results shown in the bottom row of Fig.~\ref{fig:gauss_vs_twa}, where a pronounced peak emerges at $\boldsymbol{q} = (\pi, \pi)$, accompanied only by weak features along the diagonals and near the center. Furthermore the maximum value of correlations is considerably larger in TWA compared to the truncated Gaussian result, as can be seen by comparing the color scales in Fig.~\ref{fig:gauss_vs_twa}. 

For $\Delta = 0$, the results of the truncated Gaussian approximation and TWA show closer qualitative agreement, as illustrated in Fig.~\ref{fig:gauss_vs_twa_D0}. This further supports the conclusion that the instability observed for $|\Delta| \gtrsim 1$ is driven by the dynamical enhancement of the response function, originating from quantum fluctuations, as discussed in Section~\ref{sec:longitudinal}.

	\section{RPA approximation for spin correlation functions}\label{app:RPA}

	Following the arguments of Sction~\ref{sec:Bose_stoner}, we ignore $H_4^{xy}$ in Eq.~\eqref{eq:H4} which is not important neat nesting, and consider the following Hamiltonian
	\begin{equation}
		H = \sum_{\boldsymbol{k} }\epsilon(\boldsymbol{k}) a^\dagger_{\boldsymbol{k} }a_{\boldsymbol{k}}- \frac{\Delta}{2N}\sum_{\boldsymbol{q,k,p}}J(\boldsymbol{q})\,  a^\dagger_{\boldsymbol{k+q}}a^\dagger_{\boldsymbol{p-q}}a_{\boldsymbol{p}}a_{\boldsymbol{k}},
	\end{equation}
	where $\epsilon(\boldsymbol{k})$ and $J(\boldsymbol{q})$ were given in Eqs.~\eqref{dispersion}~and~\eqref{Jz_q}, respectively. The Keldysh action for this model is given by~\cite{kamenev2023field}
	\begin{equation}
		S = S_0 + S_\mathrm{int},
	\end{equation}
	\begin{equation}
		S_0 = \sum_{s=\pm}\sum_{\boldsymbol{k} }s \int dt \, \bar{a}_{s,{\boldsymbol{k}}} \Big(i \partial_t - \epsilon(\boldsymbol{k})\Big)a_{s,\boldsymbol{k}},
	\end{equation}
	\begin{equation}
		S_\mathrm{int} = \frac{\Delta}{2N}\sum_{s=\pm} \sum_{\boldsymbol{q,k,p}} s \int dt \,  J(\boldsymbol{q})\, \bar{a}_{s,\boldsymbol{k+q}}a_{s,\boldsymbol{k}} \,\bar{a}_{s,\boldsymbol{p-q}}a_{s,\boldsymbol{p}},
	\end{equation}
	where $s=\pm$ specifies the branch of the Keldysh contour to which the fields belong. We define the density field similarly to the main text:
	\begin{equation}\label{rho_def}
		\rho_s(\boldsymbol{q},t)= \frac{1}{\sqrt{N}}\sum_{\boldsymbol{k} }\bar{a}_{s,\boldsymbol{k}}(t) a_{s,\boldsymbol{k+q}}(t),
	\end{equation}
	such that we can write $S_\mathrm{int}$ as
	\begin{multline}
		S_\mathrm{int} =\frac{\Delta}{2}\sum_{s,\boldsymbol{q}} s  \int dt \,
		J(\boldsymbol{q})\rho_s(\boldsymbol{-q},t)\rho_s(\boldsymbol{q},t)\\+\sum_{s,\boldsymbol{q}} s  \int dt \,   \varphi_s(\boldsymbol{-q},t)\bigg( \rho_s(\boldsymbol{q},t) - \frac{1}{\sqrt{N}}\sum_{\boldsymbol{k} }\bar{a}_{s,\boldsymbol{k}}(t) a_{s,\boldsymbol{k+q}}(t) \bigg),
	\end{multline}
	where $\varphi$ is Lagrange multiplier field that imposes the condition in Eq.~\eqref{rho_def}. In the next step, we apply a rotation and express fields in the classic/quantum basis define as~\cite{kamenev2023field}
	\begin{equation}
		a_c = \frac{1}{\sqrt{2}}(a_+ + a_-), \quad  a_q = \frac{1}{\sqrt{2}}(a_+ - a_-).
	\end{equation}
	In RPA, we integrate out magnon fields and keep the resulting effective action to quadratic order in $\rho$ and $\varphi$~\cite{altland2010condensed}:
    \begin{figure}[!t]
        \centering
        \includegraphics[width=0.49\linewidth]{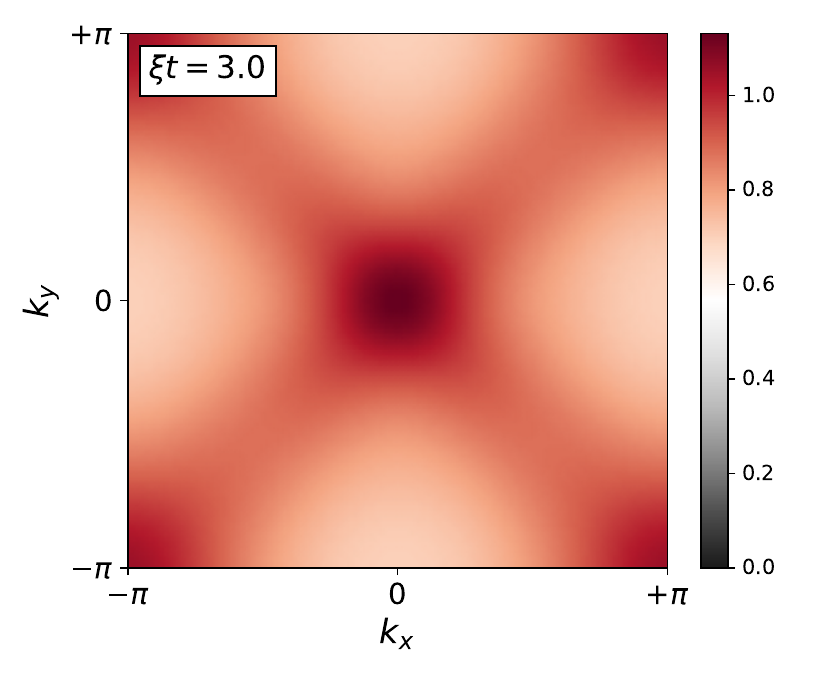}\includegraphics[width=0.49\linewidth]{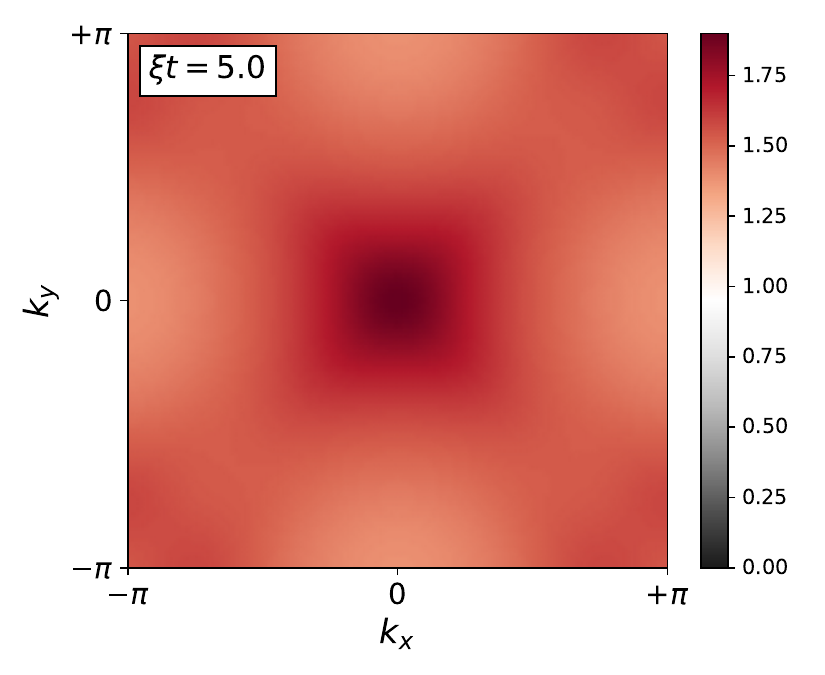}\\ \includegraphics[width=0.49\linewidth]{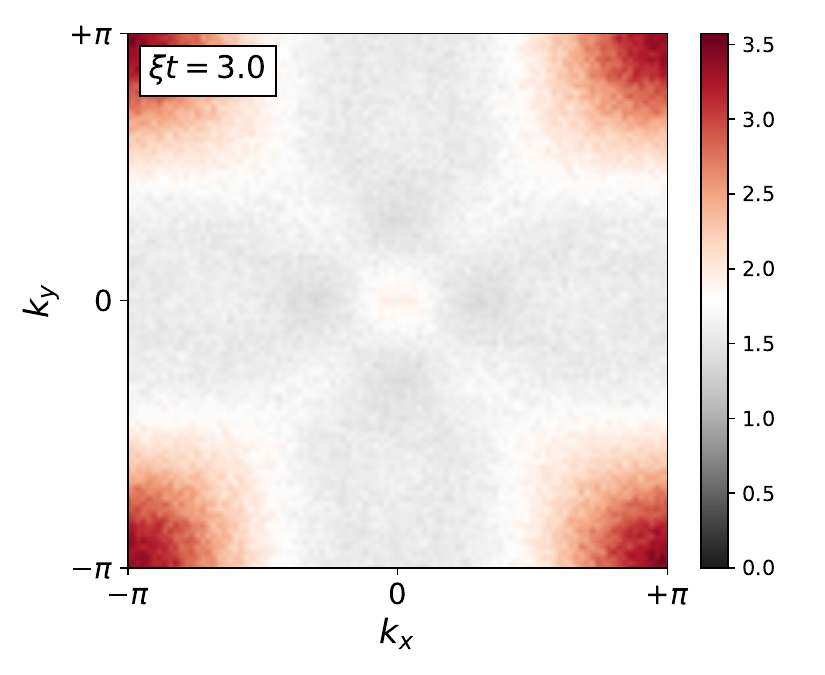} \includegraphics[width=0.49\linewidth]{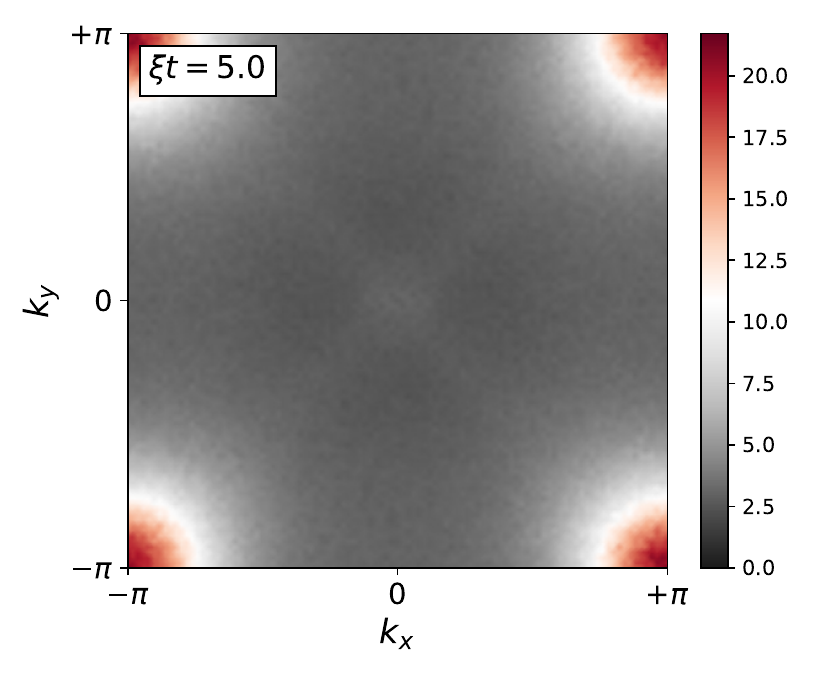}
        \caption{Longitudinal spin correlation function for $\Delta=2$, obtained from the truncated Gaussian approximation (top row) and TWA (bottom row) at two different times. The other parameters are $L=100$,  $J/h=0.2$, $\Delta=2$, $\xi/h=0.1$ and $S=5$.}
        \label{fig:gauss_vs_twa}
    \end{figure}
	\begin{equation}
		S \approx \frac12 \sum_{\boldsymbol{q}} \iint \Phi^T(\boldsymbol{-q},t) \cdot \boldsymbol{M}_{\boldsymbol{q}}(t,t')\cdot  \Phi(\boldsymbol{q},t'),
	\end{equation}
	where $\Phi(\boldsymbol{q},t)= \big(\rho_c(\boldsymbol{q},t), \varphi_c(\boldsymbol{q},t), \rho_q(\boldsymbol{q},t), \varphi_q(\boldsymbol{q},t)\big)^T$ and 
    \begin{equation}
        \boldsymbol{M}_{\boldsymbol{q}}(t,t') = \begin{pmatrix} \boldsymbol{0}_{2\times2} & M_{\boldsymbol{q}}^A(t,t') \\ & \\ M_{\boldsymbol{q}}^R(t,t') & M_{\boldsymbol{q}}^K(t,t')
			\end{pmatrix},
    \end{equation}
    with
    \begin{equation}
        \boldsymbol{M}^R_{\boldsymbol{q}}(t,t')= \big(\boldsymbol{M}^A_{\boldsymbol{q}}(t,t')\big)^\dagger = \begin{pmatrix} \Delta J(\boldsymbol{q})\delta(t-t') & \delta(t-t') \\ \delta(t-t') &  -\Pi_{\boldsymbol{q}}(t,t')
			\end{pmatrix},
    \end{equation}
    {where $\Pi_{\bm{q}}$ is the polarization function given by
    \begin{equation}
		\Pi_{\boldsymbol{q}}(\omega)=-\frac{1}{N} \sum_{\boldsymbol{k}} \frac{
			n_{\boldsymbol{k+q}} - n_{\boldsymbol{k}}
		}{\omega -\epsilon({\boldsymbol{k+q}}) + \epsilon({\boldsymbol{k}}) + i 0^+},
	\end{equation}}
    and
    \begin{equation}
        \boldsymbol{M}^K_{\boldsymbol{q}}(t,t') = \begin{pmatrix} 0 & 0 \\ 0 & +2i\tilde{C}^{zz}_{\boldsymbol{q}}(t,t')
			\end{pmatrix}
    \end{equation}
     where $\tilde{C}^{zz}_{\boldsymbol{q}}$ is the symmetric density correlation function in the truncated Gaussian approximation
	\begin{multline}
		\tilde{C}^{zz}_{\boldsymbol{q}}(t,t') =  \frac12 \Big\langle \Big\{\rho(\boldsymbol{q},t),\rho(\boldsymbol{-q},t')\Big\}\Big\rangle_\mathrm{Gauss} \\ = \frac{1}{2N} \sum_{\boldsymbol{k}} \Big(\langle a^\dagger_{\boldsymbol{k}}(t)a_{\boldsymbol{k}}(t')\rangle \langle  a_{\boldsymbol{k+q}}(t)a^\dagger_{\boldsymbol{k+q}}(t')\rangle \\ + \langle a^\dagger_{\boldsymbol{k+q}}(t')a_{\boldsymbol{k+q}}(t)\rangle \langle  a_{\boldsymbol{k}}(t')a^\dagger_{\boldsymbol{k}}(t)\rangle\Big).
	\end{multline}
     The connected part of the symmetric correlation function can be found from
    \begin{equation}
    C^{zz}_{\boldsymbol{q}}(t,t') =  \frac12 \langle \rho_c(\boldsymbol{q},t)\, \rho_c(-\boldsymbol{q},t')\rangle_c,
    \end{equation}
    which is given by the inverse of $M_{\boldsymbol{q}}$. In the frequency domain
    \begin{equation}\
        C^{zz}_{\boldsymbol{q}}(\omega)=\big[M^{-1}_{\boldsymbol{q}}(\omega)\big]_{11}=\frac{\tilde{C}^{zz}_{\boldsymbol{q}}(\omega)}{\big|1+\Delta J(\boldsymbol{q})\Pi_{\boldsymbol{q}}(\omega)\big|^2},
    \end{equation}
    which is Eq.~\eqref{C_zz_RPA_w} of the main text.

	 \begin{figure}[!t]
        \centering
        \includegraphics[width=0.49\linewidth]{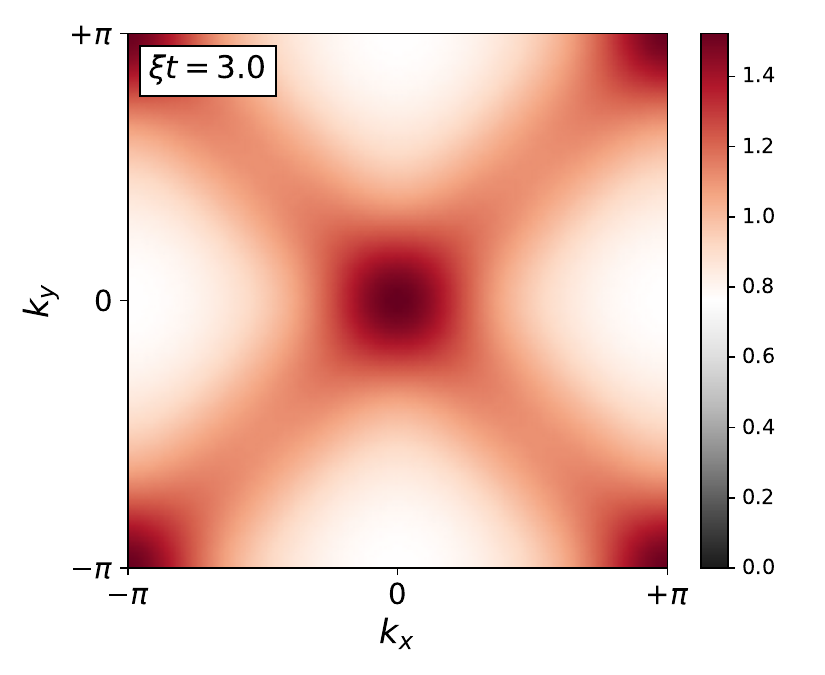}\includegraphics[width=0.49\linewidth]{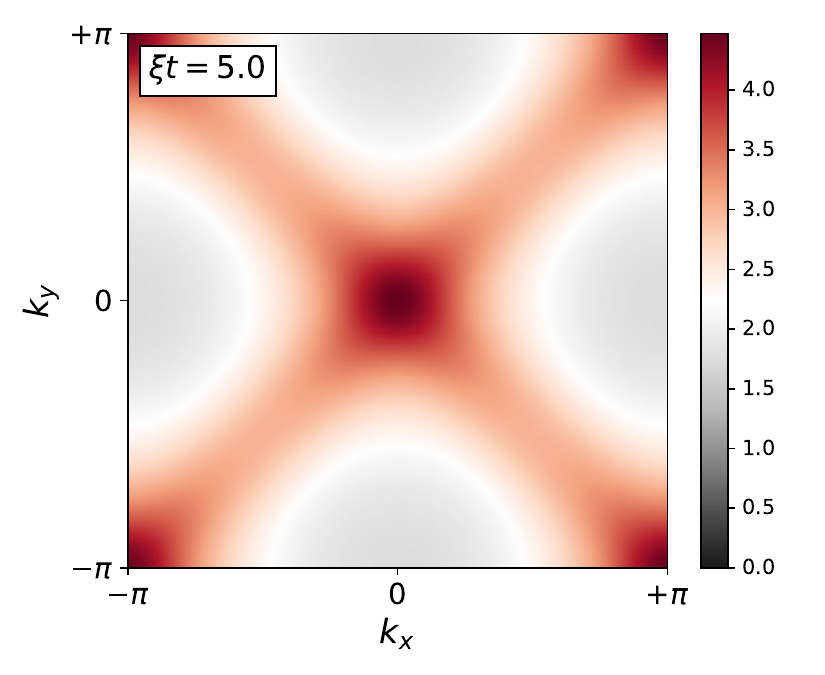}\\ \includegraphics[width=0.49\linewidth]{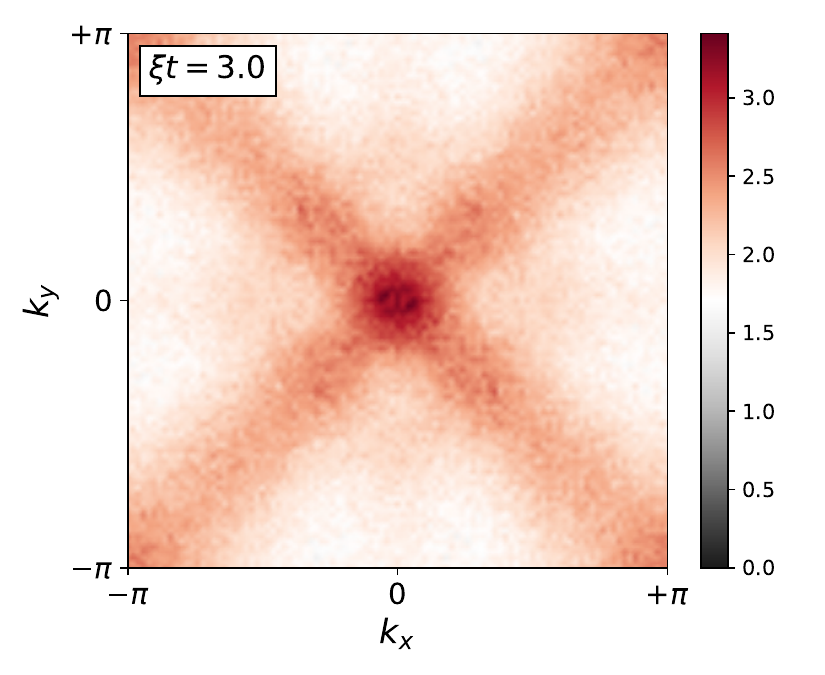} \includegraphics[width=0.49\linewidth]{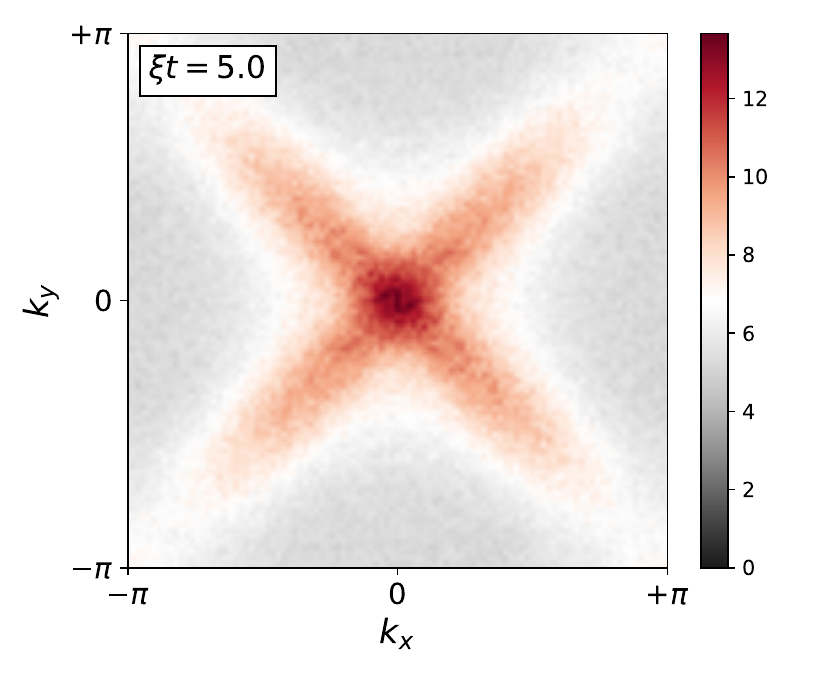}
        \caption{Longitudinal spin correlation function for $\Delta=0$, obtained from the truncated Gaussian approximation (top row) and TWA (bottom row) at two different times. The other parameters are $L=100$,  $J/h=0.2$, $\Delta=2$, $\xi/h=0.1$ and $S=5$.}
        \label{fig:gauss_vs_twa_D0}
    \end{figure}

\section{Chiral transformation}\label{app:chiral}

	We show that for the XX Hamiltonian ($\Delta=0$ in Eq.~\eqref{H_XXZ}), there is a mapping between the magnon populations generated by drives with opposite detunings:
    \begin{equation}\label{eq:app_delta_opposite}
        \omega_\pm = \omega_\mathrm{nest} \pm \delta\omega.
    \end{equation}
	We discard the counter-rotating terms in the driving Hamiltonian, and go to a frame rotating with the driving frequency. The Hamiltonians for Eq.~\eqref{eq:app_delta_opposite} in the rotating frame are given by
    \begin{equation}
        H_\mathrm{\pm} = \pm H_{Z} + H_{XX},
    \end{equation}
    where we have defined the detuning Hamiltonian as
    \begin{equation}
        H_Z = -\frac{\delta \omega}{2} \sum_i S^z_i,
    \end{equation}
    together with the nonlinear part
    \begin{equation}
        H_{XX} =  - \frac{J}{4S}\sum_{\expval{ij}} \Big(S^x_i S^x_j + S^y_i S^y_j\Big) + \frac{\xi}{8S} \sum_i \Big[(S^+_i)^2+(S^-_i)^2\Big].
    \end{equation}
    As was introduced in Eq.~\eqref{eq:chiral_trans} of the main text, we consider the following unitary transformation:
    \begin{equation}
        U =  \exp \Big[-\frac{i\pi}{2}\sum_{j_x,j_y} (-1)^{j_x+j_y} S^z_j \Big].
    \end{equation}
    $U$ transforms spin operators on the even and odd sub-lattices differently:
    \begin{align}
        US^+_j U^\dagger &= \begin{cases}
            -i S^+_j & j \in \mathrm{even},\\ +i S^+_j & j \in \mathrm{odd},
        \end{cases}\\
        U S^z_j U^\dagger &= S^z_j.
    \end{align} 

    As a result, $U$ transforms $H_{Z}$ and $H_{XX}$ according to
    \begin{equation}
        UH_{Z}U^\dagger = H_Z, \qquad U H_{XX}U^\dagger=-H_{XX},
    \end{equation}
	which leads to the following mapping between the oppositely-detuned cases
    \begin{equation}\label{eq:app_chiral_map_Hxx}
        U H_{\pm} U^\dagger = - H_{\mp}.
    \end{equation}
	Therefore, the spectra of $H_+$ and $H_-$ are opposite. Moreover, $U$ shifts the momentum of spin-waves by $(\pi,\pi)$, as was shown in Eq.~\eqref{eq:chiral_S}. In summary, we have shown that for $\Delta=0$:
    \begin{enumerate}
        \item For each excited magnon of $H_+$, labeled by $\ket{\lambda,\boldsymbol{k}_+}$ with energy $E_\lambda$, $H_-$ has a dual magnon excitation $\ket{\lambda,\boldsymbol{k}_-}$ with opposite energy $-E_\lambda$.
        \item These dual eigenstates are related by a momentum shift
        \begin{equation}
            \boldsymbol{k}_+ = \boldsymbol{k}_- + (\pi,\pi).
        \end{equation}
    \end{enumerate}
    Fig.~\ref{fig:duals} shows an example, where the magnon profile of two oppositely detuned drives are clearly connected by the vector $(\pi,\pi)$. For $\Delta\neq 0$, the mapping in Eq.~\eqref{eq:app_chiral_map_Hxx} is not valid anymore since the $\Delta$ term is even under $U$. Furthermore, we cannot interpret the coefficient in $H_Z$ as the detuning from the nesting frequency, since the latter now has a contribution from $\Delta$ (Eq.~\eqref{omega_nest}).

\begin{figure}[!t]
        \centering
        \hspace{-5pt}\includegraphics[width=.98\linewidth]{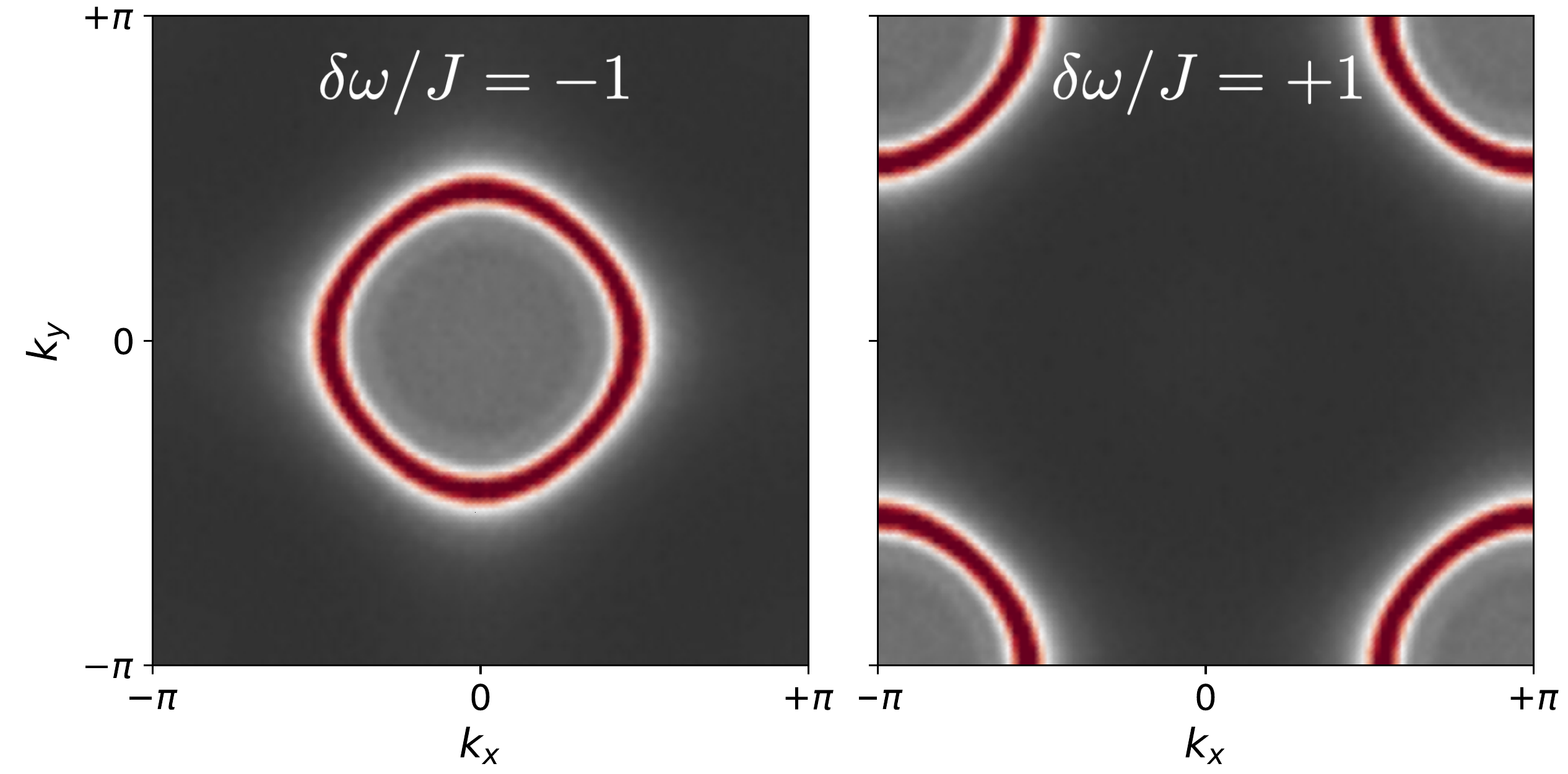}
        \caption{Spin-wave population of the XX Hamiltonian for two drives with opposite detunings from the nesting frequency, at time $\xi t = 7$ after starting the drive. The populations can be mapped to each other through a momentum shift of $(\pi,\pi)$. Other parameters are $h/J = 1.25$, $\xi/J=0.25$, $S=5$, and $N=100\times100$.}
        \label{fig:duals}
    \end{figure}
    
    We remark that the above duality is approximate, as we had to neglected the counter-rotating terms in $H_\mathrm{drive}$ for our argument. However, the counter-rotating terms are unimportant over long timescales, which scale exponentially with $\omega/\xi$, as will be discussed below.

	\section{Heating}\label{app:heat}
	
			\begin{figure}[!t]
		\centering
		\subfloat[\label{fig:heating_xxz}]{\includegraphics[height=0.72\linewidth,trim={0 .3cm 0.2cm 0.25cm},clip]{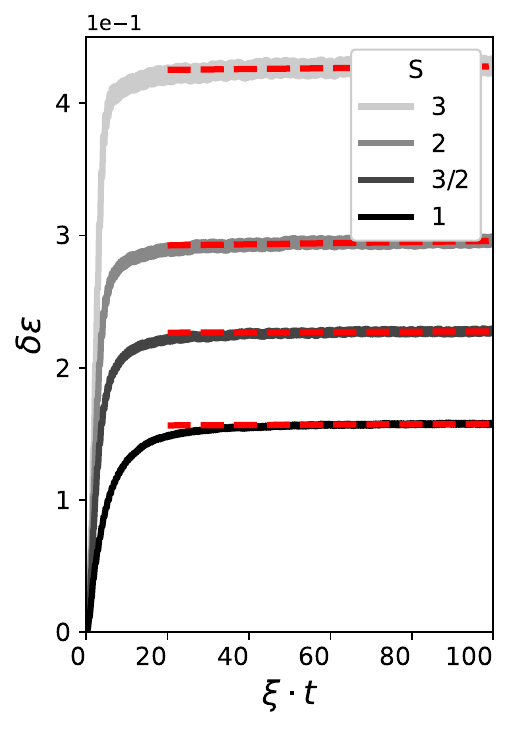}}\hspace{3pt}
        \subfloat[\label{fig:heating_xx}]{\includegraphics[height=0.72\linewidth,trim={.9cm .3cm 0.2cm 0.25cm},clip]{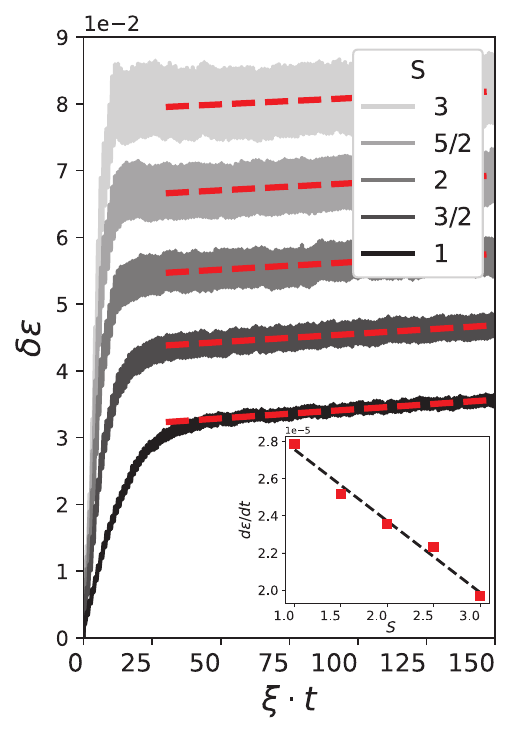}}
		\caption{Evolution of energy density under driving for different spin sizes in a system of size $20\times20$. (a) For the XXZ model with $J/h=0.2$, $\Delta=-2$, $\omega/h=1.2$, and $\xi/h=0.2$. (b) For the XX model with $J/h=0.8$, $\Delta=0$, $\omega/h=0.6$, and $\xi/h=0.2$. In both cases the energy quickly grows at early times, before reaching a prethermal plateau which is characterized by a slow absorption of energy at a linear rate which decreases with $S$. Inset: long-time heating rates for the data in panel (b), showing slower heating for larger spins.}
	\end{figure}
	
	Heating is a generic feature of driven classical and quantum many-body systems, which leads the system towards a featureless,  maximal entropy state. All of the presented results in this work regarding instabilities are, in principle, valid for a finite window of time, after which heating washes out these features. Therefore, it is necessary to characterize heating effects in our system, and to ascertain that the non-equilibrium state survives long enough before being destroyed by heating.
	
	In Fig.~\ref{fig:heating_xxz}, we have shown the evolution of energy density $\delta\varepsilon = (\bra{\psi(t)} H \ket{\psi(t)}-\bra{\psi_0} H \ket{\psi_0})/N$, where $H$ was given by Eq.~\eqref{H_XXZ}, for different spin sizes after switching on the drive. For all values of $S$, we observe an initial period of energy growth, during which the resonant drive injects magnons (which carry energy) into the system at a finite rate, consistent with our expectations for the driven model in Eqs.~\eqref{H_XXZ}~and~\eqref{H_drive}. Followed by the early quick heating, there is a crossover to a regime of slow (pre)thermalization, with slower heating for larger spin sizes. In the parameter regime relevant for magnon nesting ($|\Delta| \gtrsim 1$), the heating rate is very small across all spin sizes. Combined with the stochastic nature of our approximation, which requires averaging over many trajectories to accurately capture long-time behavior, this made it challenging to precisely determine the dependence of the heating rate on $S$. However, we were able to access the late-time heating dynamics more reliably in the $\Delta = 0$ case, where a similar two-stage behavior emerges (Fig.~\ref{fig:heating_xx}). In this regime, the late-time heating rate could be determined with greater accuracy and was found to decrease with increasing spin size.
	
	The origin of the two-stage relaxation profile can be explained by looking at the system in the reference frame rotating with $\omega$, in which dynamics is captured by the Hamiltonian $H_\mathrm{rot}(t)=H_1 + H_2$. $H_1$ is the non-oscillatory part given by
	\begin{multline}\label{H_rot_1}
		H_1= \tilde{h} \sum_i S^z_i - \frac{J}{4S}\sum_{\expval{ij}}\Big(S^x_i S^x_j + S^y_i S^y_j  + \Delta \, S^z_i S^z_j \Big) \\+ \frac{\xi}{8S}\Theta(t) \sum_i \Big[(S^+_i)^2+(S^-_i)^2\Big],
	\end{multline}
    where $\tilde{h}=h-\omega/2$ is the shifted magnetic field. $H_2(t)$ is the counter-rotating part
	\begin{equation}\label{H_rot_2}
		H_2= \frac{\xi}{8S} \Theta(t) \sum_i \Big[e^{2i\omega t}(S^+_i)^2+e^{-2i\omega t}(S^-_i)^2\Big].
	\end{equation}
    For our regimes of interest, terms in $H_2$ rotate fast and can be ignored for timescales which grow at least exponentially in $\omega/\xi$~\cite{abanin2017rigorous}. The remaining part of the Hamiltonian ($H_1$) describes a quench, in which a finite amount of energy is injected into the system after a sudden change of $\xi$ from zero to a finite value, resulting in the increase of the energy density at early times. In the absence of $H_2$, the energy would reach a plateau, corresponding to a thermal state of $H_1$. However, in our case, the rotating term $H_2$ heats up the system at longer times, and adds a finite slope to the plateau. The heating is slow because, if we treat magnons as free particles, the driving by $H_2$ is off-resonant with respect to the energy of magnon pairs. As a result, the energy exchange occurs only due to magnon nonlinearities.

	Last, we discuss how larger spin sizes suppress heating at long times. As was discussed above, the slow heating is generated by $H_2$, whose frequencies are far off-resonant with respect to the spectrum of free magnons. If the energy levels of $H_1$ are given by $\ket{n}$, then $H_2$ can create magnon pairs if the spectral density 
	\begin{equation}
		\mathcal{A}_\mathrm{pair}(\omega)= \sum_{n,m} \rho_{mm}\Big|\bra{n} \sum_{\boldsymbol{k}}a^\dagger_{{\boldsymbol{k}}}a^\dagger_{-{\boldsymbol{k}}} \ket{m}\Big|^2 \delta(2\omega - E_{nm}),
	\end{equation}
	is finite. In the above expression, $\rho_{mm}$ is the density matrix and $E_{nm}=E_n-E_m$. For free magnons  we always have $E_{nm}=\tilde{\epsilon}({\boldsymbol{k}})+\tilde{\epsilon}(-{\boldsymbol{k}})$, where $\tilde{\epsilon}({\boldsymbol{k}})$ is the magnon dispersion of $H_1$. Therefore, $H_2$ cannot induce any transitions when $\omega > \max(\tilde{\epsilon}({\boldsymbol{k}}))$. However, including magnon nonlinearities changes this picture, and the energy of a multi-magnon state is not equal to the sum of the energies of an equal number of free magnons. An expansion of the bosonic representation (Eq.~\eqref{HP}) in powers of $S$ shows that nonlinearities appear at $\mathcal{O}(S^{-1})$ in the Hamiltonian. As a result, for larger values of $S$ the magnon spectrum deviates less from the non-interacting limit, reducing the possibility of interaction-induced resonances in the many-body spectrum, and consequently, the efficiency of $H_2$ to heat the system.

	\bibliography{Refs}

\end{document}